\documentclass[11pt,a4paper]{article}
\pdfoutput=1
\usepackage{jcappub}
\usepackage{geometry}
\usepackage{subfigure} 
\usepackage{hyperref}
\usepackage{float}

\makeatletter 
\DeclareRobustCommand*\textsubscript[1]{%
  \@textsubscript{\selectfont#1}}
\def\@textsubscript#1{%
  {\m@th\ensuremath{_{\mbox{\fontsize\sf@size\z@#1}}}}}
\makeatother

\newcommand\be{\begin{equation}} 
\newcommand\ee{\end{equation}} 
\newcommand\bea{\begin{eqnarray}} 
\newcommand\eea{\end{eqnarray}}

\newcommand{\snana}{{\tt SNANA}}

\newcommand{\mlcs}{{\tt MLCS2k2}}
\newcommand{\salt}{{\tt SALT2}}
\newcommand{\saltm}{{\tt SALT2mu}}
\newcommand{\fpmlcs}{$f_{p}$MLCS}
\newcommand{\fpsalt}{$f_{p}$SALT}

\newcommand{\aj}{AJ}
\newcommand{\apj}{ApJ}

\newcommand{\pasp}{PASP}
\newcommand{\mnras}{MNRAS}

\title{Type Ia Supernovae Selection
and \\ Forecast of Cosmology Constraints \\for the Dark Energy Survey} 
\author[a,b]{Eda Gjergo,}
\author[c]{Jefferson Duggan,}
\author[c,a]{John D. Cunningham,}
\author[a]{Steve Kuhlmann,} 
\author[a]{Rahul Biswas,}
\author[a]{Eve Kovacs,}
\author[a]{\\ Joseph P. Bernstein,}
\author[a]{Harold Spinka}

\affiliation[a]{Argonne National Laboratory \\
9700 South Cass Avenue, Lemont, IL 60439, USA}
\affiliation[b]{Illinois Institute of Technology \\
Applied Mathematics Office, E1 Building 10 West 32nd Street, Chicago, IL 60616}
\affiliation[c]{Department of Physics, Loyola University Chicago \\
1032 W. Sheridan Road, Chicago, IL 60660}

\emailAdd{eda.gjergo@gmail.com}
\emailAdd{jnaggud@gmail.com}
\emailAdd{jcunni6@luc.edu}
\emailAdd{kuhlmann@anl.gov}
\emailAdd{kovacs@anl.gov}
\emailAdd{rbiswas4@gmail.com}

\abstract{We present the results of a study of selection criteria to
identify Type Ia supernovae photometrically in a simulated mixed
sample of Type Ia supernovae and core collapse supernovae. The simulated
sample is a mockup of the expected results of the
Dark Energy Survey.  
Fits to the \mlcs\ and \salt\ Type Ia supernova models 
are compared and used to help separate the 
Type Ia supernovae from the core collapse sample.  The Dark Energy 
Task Force Figure of Merit (modified to include core collapse supernovae
systematics) is used to discriminate among the various selection 
criteria. This study of varying
selection cuts for Type Ia supernova candidates is the first to 
evaluate core collapse contamination using the Figure of Merit.
Different factors that contribute to the Figure of
Merit are detailed.  With our analysis methods, 
both \salt\ and \mlcs\ Figures of Merit improve
with tighter selection cuts and higher
purities, peaking at 98\% purity.} 


\date{\today}

\begin{document}
\maketitle
\flushbottom

\section{Motivation}\label{sec:motivation}
In the next decade, the number of detected Type Ia supernovae (SNIa) will increase 
dramatically \cite{bern11,lsst},  surpassing the resources available for spectroscopic 
confirmation of each supernova (SN).  This has produced an increased interest 
in the photometric identification of SNIa in samples including significant 
numbers of core collapse supernovae (SNcc).  In order to improve the 
contraints on the accelerated expansion of the universe,  discovered with
SNIa in the late 1990's~\cite{rie98,per99}, photometric typing of 
supernovae (SNe) must be 
very robust.  Two recent studies of simulated SNe have approached
the subject of SN photometric identification in different ways: 1) the first, 
Kessler~{\it{et~al.}}~(2010)~\cite{SNchall},
compared a wide variety of photometric-typing algorithms,  but did 
not evaluate the impact on cosmology constraints, 2) the second, 
Bernstein~{\it{et~al.}}~(2011)~\cite{bern11}
studied the impact on cosmology in detail, but only used one
photometric-typing algorithm (\mlcs\ fit quality cuts).  The analysis presented in this
paper is a follow-up to Bernstein~{\it{et~al.}}, and incorporates several  
new features in the analysis.

Using tight signal-to-noise ratio (SNR) cuts and a SNIa fit quality cut 
(\mlcs\ model~\cite{jha07}),
Ref.~\cite{bern11} achieved a purity (SNIa/Total) above 95\%. The remaining SNcc had 
 a negligible effect on cosmology. This was achieved for the case where the
redshifts of the SNe were assumed to be measured accurately in a spectroscopic
follow-up of the host galaxies. 
In this article, we follow a similar approach but extend the 
analysis by studying the effects of 
relaxing the SNR cuts and including the \salt\ SNIa model fit
quality.  We use data samples simulated for 
the 10-field hybrid footprint of the
Dark Energy Survey\footnote{http://www.darkenergysurvey.org} (DES),
performed with the \snana\ package as 
in Ref.~\cite{bern11}.
We have updated the SNcc simulation inputs to reflect 
improved knowledge of their relative fractions and 
brightnesses (see section $\S$\ref{sec:siminput}).
We present four distinct sets of SNR cuts for both the \mlcs\ and \salt\ 
models (using the \saltm\ procedure in Ref.~\cite{mar11} to 
obtain distance moduli for the \salt\ 
model).\footnote{The \salt\ simulations used $\alpha=0.135$ and 
$\beta=3.19$.  The \saltm\ evaluation of the distance 
modulus also used the same fixed values of $\alpha$ and 
$\beta$. The fitting of $\alpha$ and $\beta$, in the 
presence of significant SNcc contamination, is beyond
the scope of this paper.}
The quantity SNRMAX is defined to be the SNR at 
the measured epoch of maximum signal-to-noise in each 
of the four DES broadband filters used in the supernova analysis. 
Within a single survey strategy in terms of average observing conditions
and cadence, this quantity may be used as a rough proxy for how well the 
supernova light curve was measured. 
Our goal is to find the purity levels that optimize
the Dark Energy Task Force (DETF)~\cite{alb06} Figure of Merit (FoM).
One obvious question is: What is a significant level of 
change in FoM?
Our goal is to have the uncertainty in FoM due to the SNcc
sample to be much smaller than the largest uncertainty in Ref.~\cite{bern11}, which
was due to the filter zeropoint uncertainty.  
The filter zeropoint uncertainty caused a 70 unit reduction in the FoM (30\%).
Therefore, we consider changes of $>$10\%
to be significant in our analysis.  We are not considering
the entire suite of systematic uncertainties in this analysis,  only the 
impact of photometric typing and selection cuts for the mixed SNe sample.

The outline of the paper is as follows.  We present the changes
in the simulation of the SNcc sample in $\S$\ref{sec:siminput}.
Our variety of SNR selection criteria,  the SNIa models we are 
using, and the resulting purities and efficiencies are presented in $\S$\ref{sec:selection}.  
The DETF Figure of Merit calculation, some
relevant factors and examples, and the final results are 
presented in $\S$\ref{sec:detffom}.  We discuss the
results in $\S$\ref{sec:discuss} and include more 
details of the new simulation inputs in Appendix~A.
Finally, we include supplementary figures in Appendix~B.

\section{Supernova Sample Simulations}\label{sec:siminput}
The \snana\ package~\cite{snana} is used to simulate the light curves
of the 5-year SNIa and SNcc samples for the DES.  The simulations are 
very similar to those in Ref.~\cite{bern11} but 
have been updated and improved with more recent information.  
The list of changes since Ref.~\cite{bern11} are:

\begin{itemize}
 \item The \snana\ version was updated to v9\_89b from v8\_37.  Our model choices 
 were MLCS2k2.v007 and SALT2.LAMOPEN. 
 \item Four more SNcc templates (2 Ib/c and 2 IIP) are added to the 40 templates
used in Ref.~\cite{bern11}. The templates are from 
       the Supernova Photometric Classification Challenge~\cite{SNchall}.
 \item The results of Li {\it{et al.}}~\cite{Li11} are now used for 
   the relative fractions of the SNcc sample, instead of those of
   Smartt {\it{et al.}}~\cite{sma09},  due to a better analysis of 
   the sample completeness.
 \item The Li {\it{et al.}}~\cite{Li11} results are now used for 
   the absolute brightnesses of the SNcc sample, instead of those
   of Richardson {\it et al.}~\cite{ric02}.  This is also due to the 
   better analysis of sample completeness. 
 \item We now use separate relative fractions for SN Types Ib and Ic, as 
   well as different average brightnesses based on Li {\it{et al.}}.
   (Type II SNe already had separate relative fractions in \cite{bern11}.)
 \item Since the Li {\it{et al.}} sample is complete, and the absolute 
  brightnesses are not corrected for dust extinction, our simulation does 
  not include dust extinction applied to the SNcc sample. 
 \item The widths of the absolute brightness distributions for each type of 
  SNcc template are matched to the measured widths from Li {\it{et al.}} 
\end{itemize}

The details of these changes are available in Appendix~A.  The changes 
mostly cancel each other in terms of the overall purity of the sample
with Ref.~\cite{bern11} cuts.  The relative mixture of the 
SNcc sample passing cuts is more uniform, however, with less dominance
from the Type Ibc SNe.

\section{Selection Criteria and Type Ia SN models}\label{sec:selection}

\subsection{Supernova Sample Signal-to-Noise Cuts}\label{sec:cuts}

As mentioned previously, our SNR is defined at 
the measured epoch of maximum SNR in each filter (SNRMAX). 
We present four distinct sets of SNR cuts for both the \mlcs\ and \salt\ models.
For simplicity, we define the symbols used in the rest of the paper for 
these four sets of cuts in Tab.~\ref{tab:symbols}.
 
\begin{figure}[t]
\begin{center}
\includegraphics[width=0.49\columnwidth]{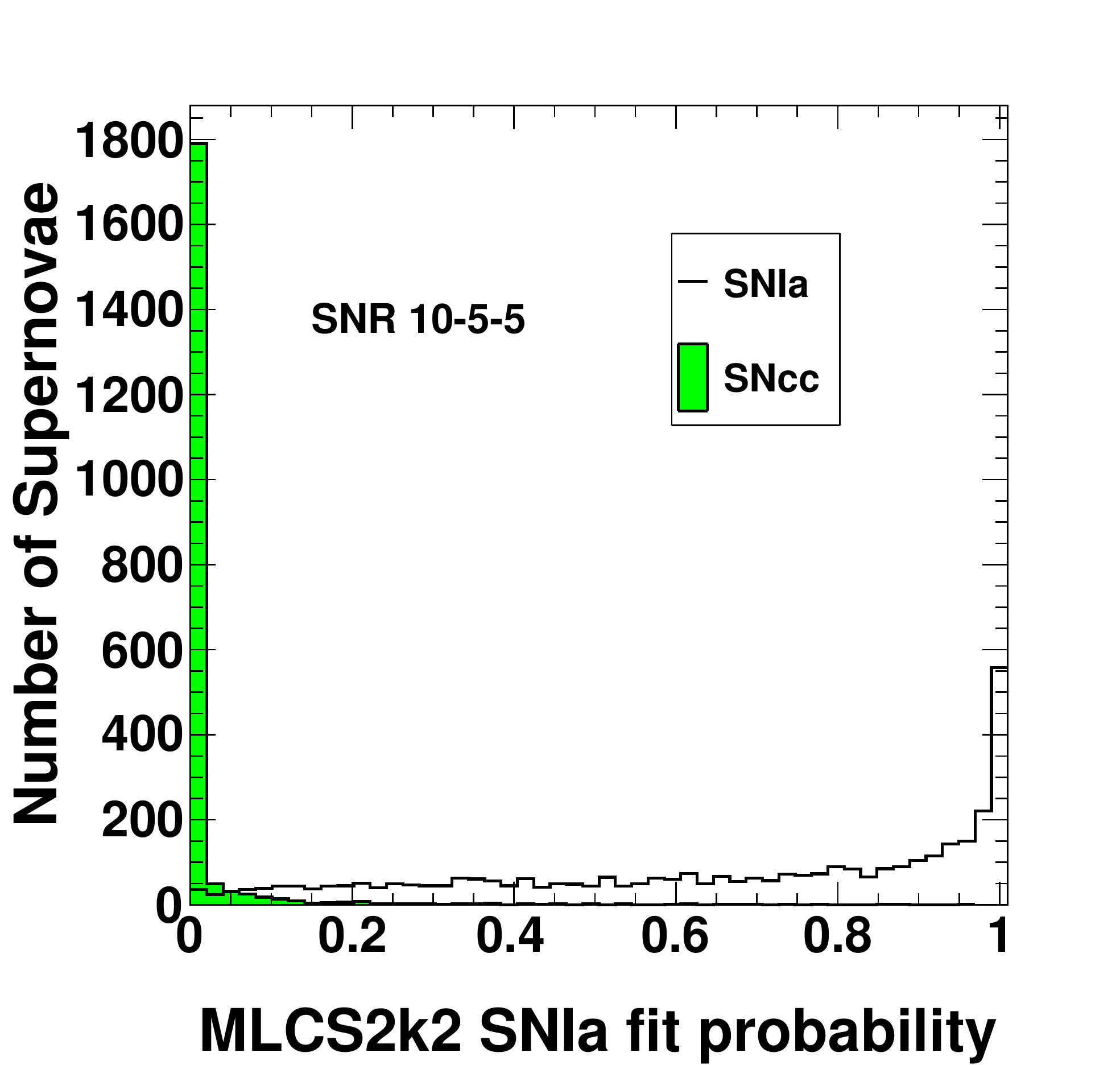}
\includegraphics[width=0.49\columnwidth]{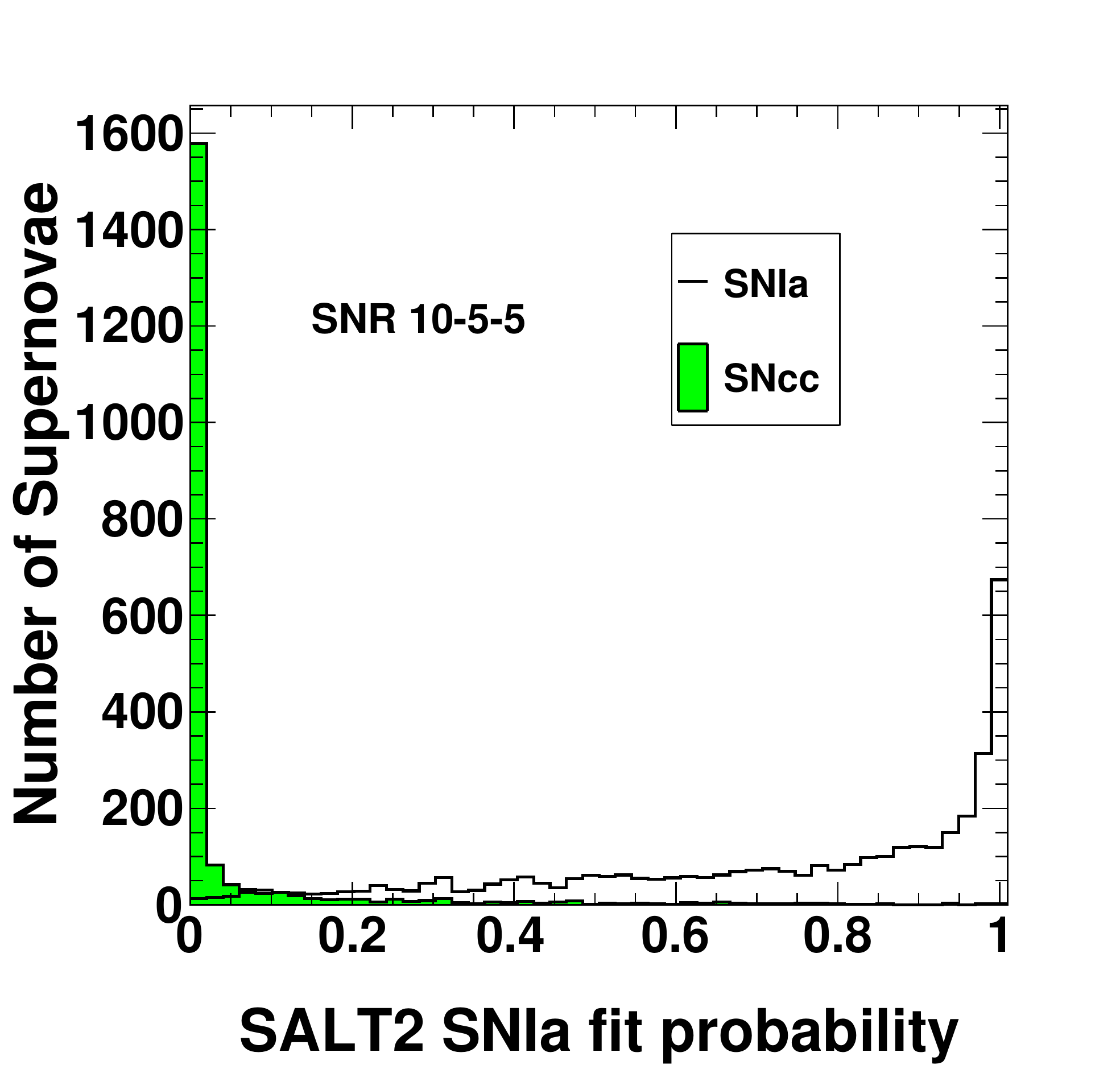}
\includegraphics[width=0.49\columnwidth]{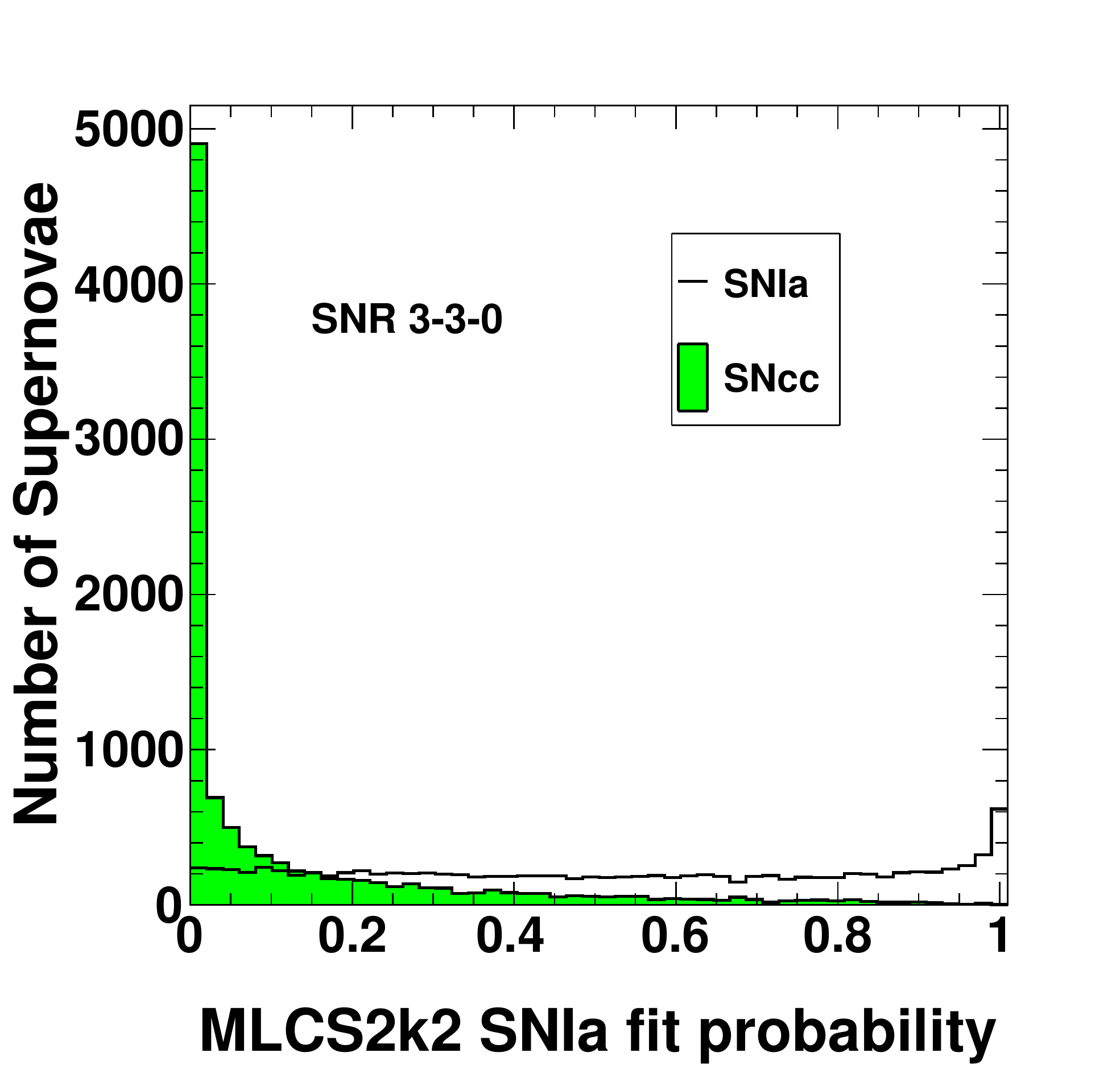}
\includegraphics[width=0.49\columnwidth]{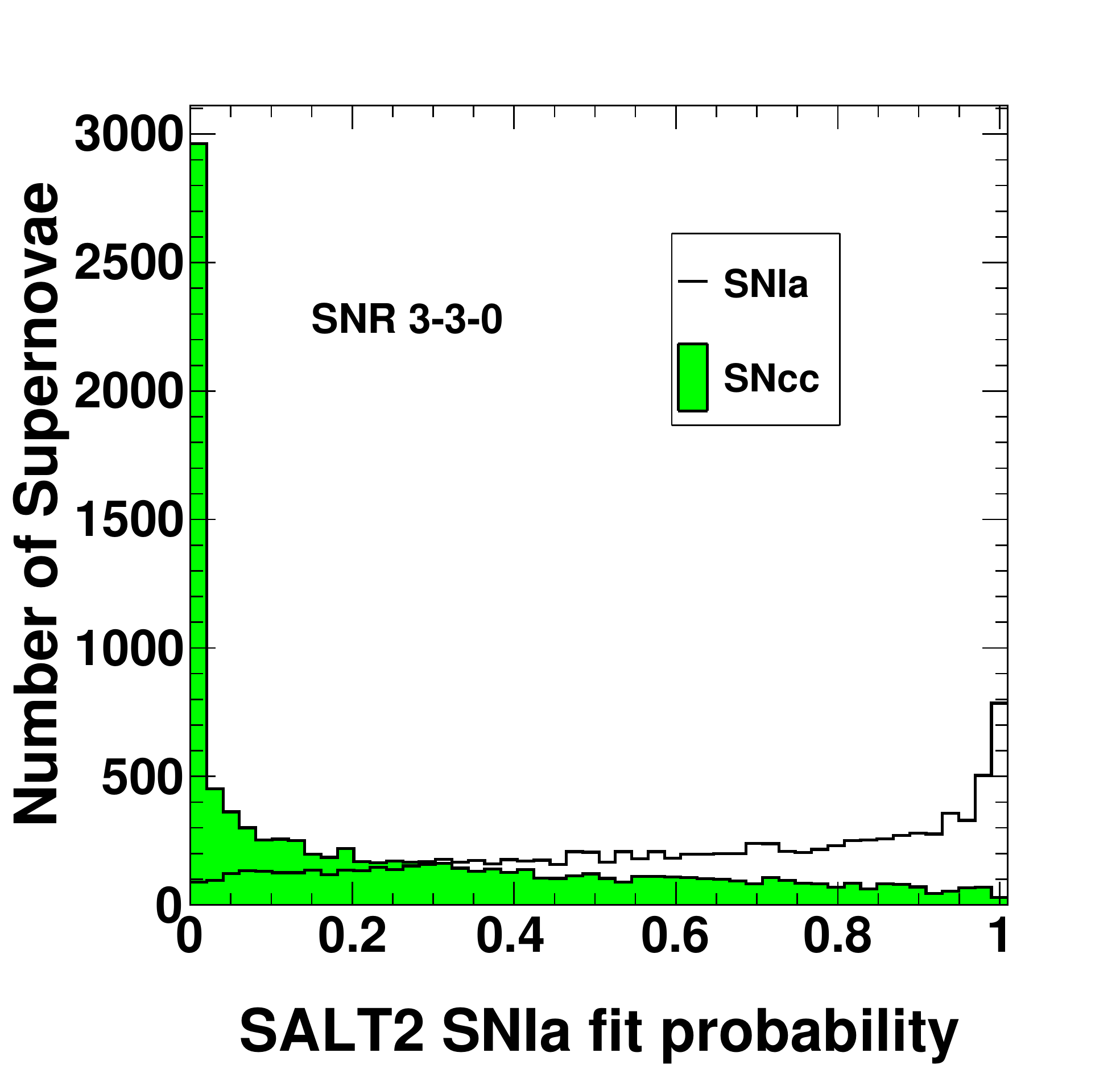}
\caption{The \mlcs\ fit probabilities ({\it left panels}) are plotted for the 
SNIa and SNcc samples. The top left panel is for the tightest 
SNR cuts, SNR-10-5-5,  while the bottom left panel is for the loosest SNR 
cuts, SNR-3-3-0.  The corresponding \salt\ fit probabilities are shown in the right panels, 
with the tightest cuts on top and the loosest cuts on bottom.}
\label{fig:fp_compare}
\end{center}
\end{figure}

\begin{table}[h]
\begin{center}
\small
\begin{tabular}{|c|c|} 
\hline
Cuts & Symbol \\\hline \hline
2 filters with SNRMAX $\geq 3$ & SNR-3-3-0 \\
2 filters with SNRMAX $\geq 5$ & SNR-5-5-0 \\
3 filters with SNRMAX $\geq 5$ & SNR-5-5-5 \\
1 filter SNRMAX $\geq 10$, 2 more filters SNRMAX $\geq 5$ & SNR-10-5-5\\ 
\hline
\end{tabular}
\caption{These are the definitions of the signal-to-noise cut 
symbols used throughout this work. For the first two cuts listed in the table,
we removed SNe for which  the third filter was less than zero. For the remainder of the paper
when we refer to the ``tightest'' and ``loosest'' SNR cuts we 
mean SNR-10-5-5 and SNR-3-3-0 respectively.}\label{tab:symbols}
\end{center}
\end{table}

\subsection{Type Ia Model Fit Probabilities}\label{sec:Iamodels}
Core collapse SNe light curves fit to a SNIa light curve model might be 
expected to have bad fit qualities, and this was demonstrated 
in Ref.~\cite{bern11}. Motivated by this, 
we reject SN candidates which have deviations from the best fit light curve 
model (whether \mlcs~or \salt ) that are statistically large compared to the errors. This is quantified
in terms of fit probabilities\footnote{If the observed 
deviations from the best fit model were due to Gaussian fluctuations 
compatible with the reported errors on observations, this is the 
probability of the $\chi^2$ being larger than the 
observed $\chi^2$ for the number of degrees of freedom in the light 
curve fit.} 
obtained from the light-curve $\chi^2$ and the
number of degrees of freedom. 
Figure~\ref{fig:fp_compare} shows the results of the fit probabilities for
both models, and for our tightest and loosest cuts.
It is evident that the \salt\ model has larger fit probabilities
for the SNcc sample and hence we obtain lower purities for 
\salt\ compared to those of \mlcs.  This is most likely 
due to the use of tight dust extinction priors used in
the \mlcs\ fits~\cite{bern11}.
But as described in Ref.~\cite{bern11}, these \mlcs\ priors lead to 
additional SNe color systematics not present in \salt\ fits.

\subsection{Purities and Efficiencies}\label{sec:purities}
\begin{figure}[ht]
\begin{center}
\includegraphics[width=0.49\columnwidth]{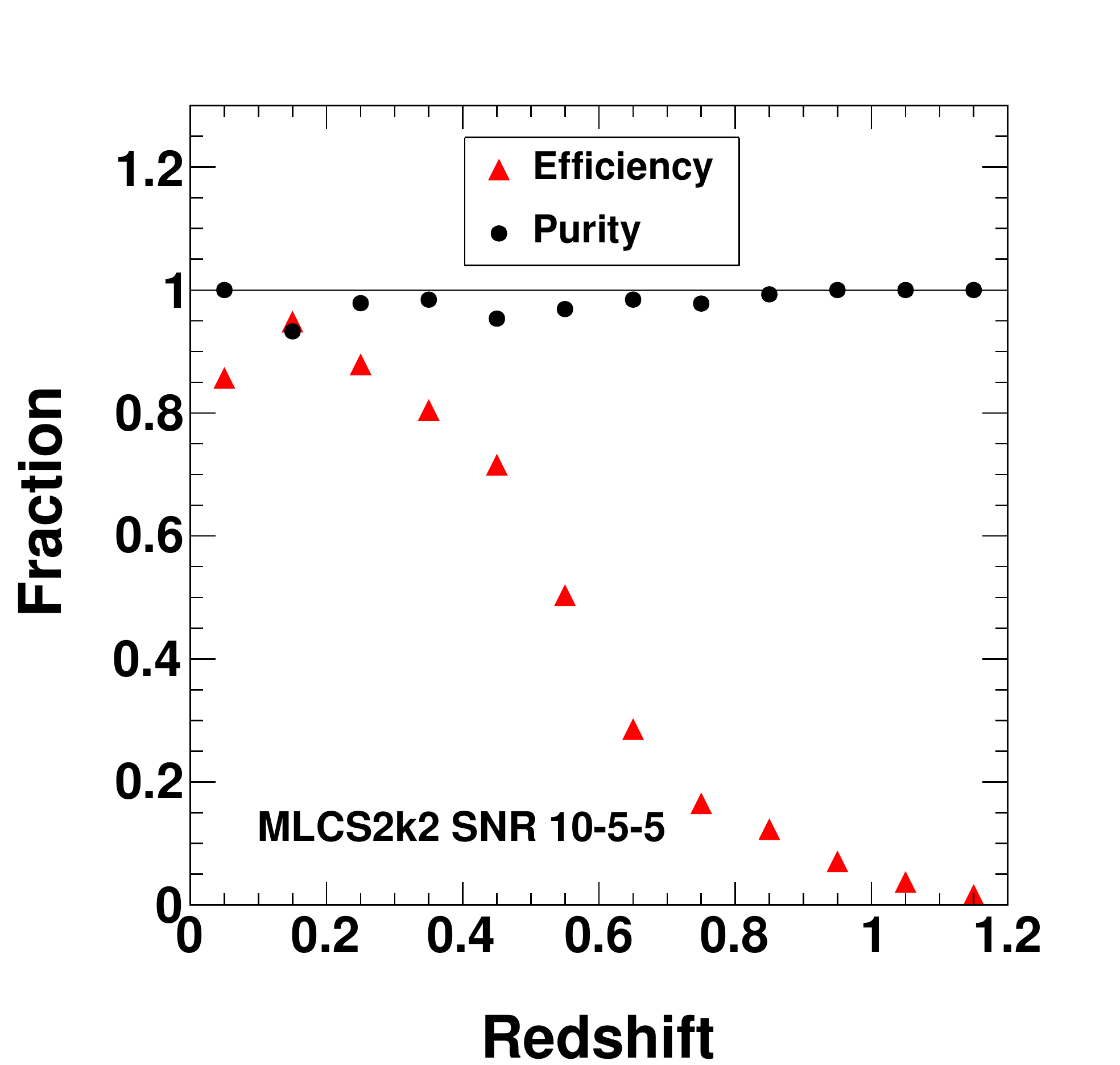}
\includegraphics[width=0.49\columnwidth]{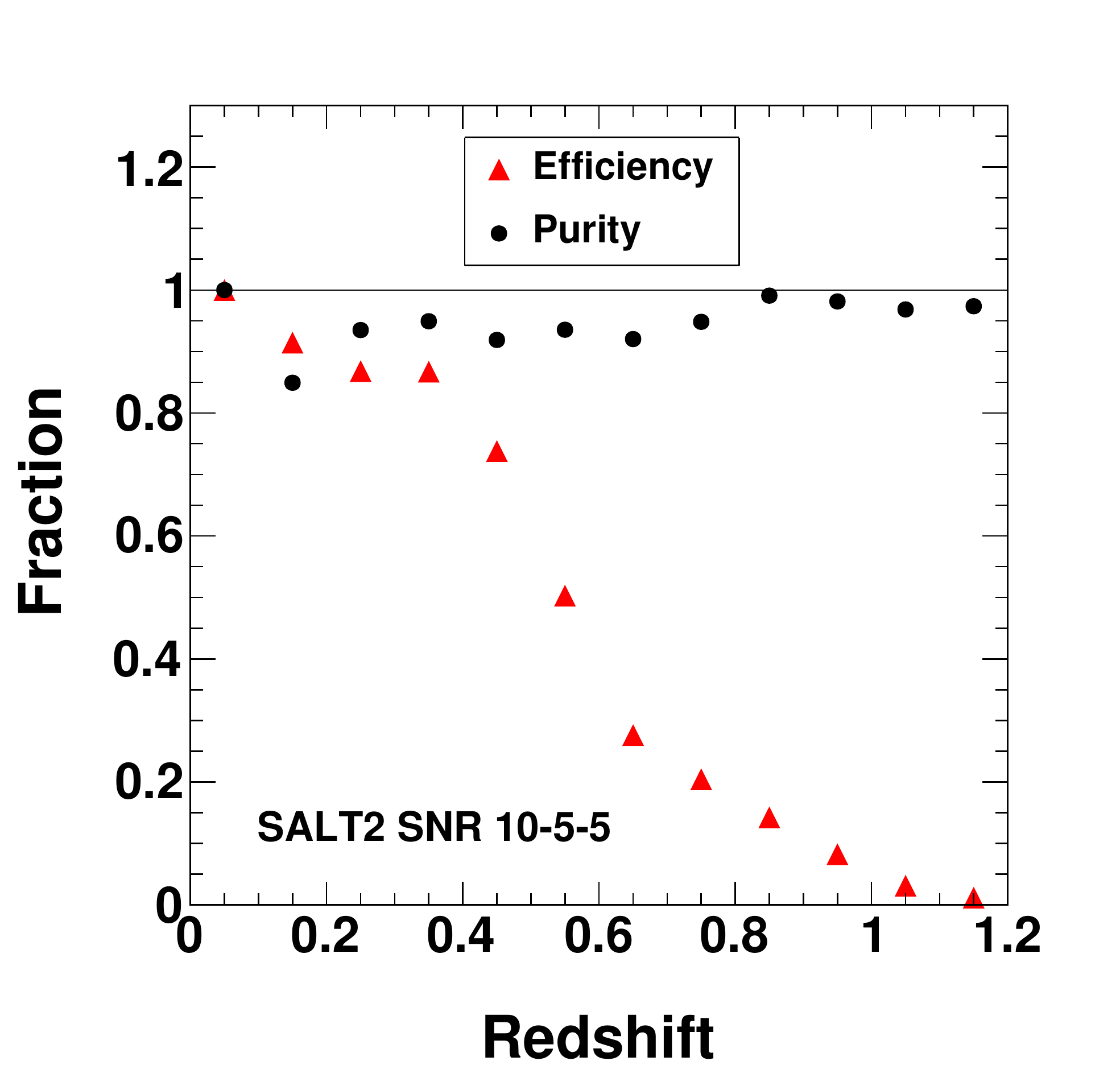}
\includegraphics[width=0.49\columnwidth]{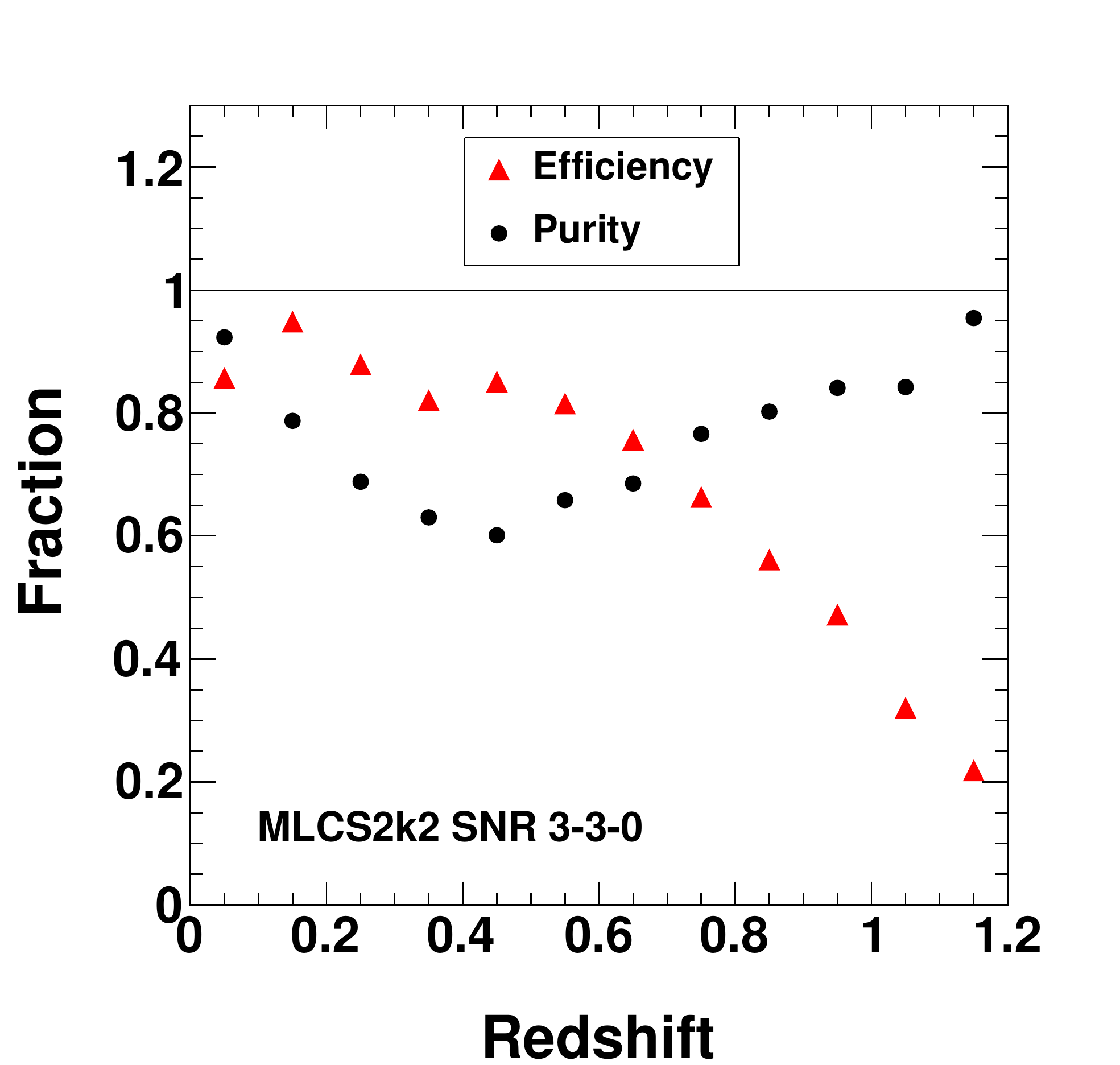}
\includegraphics[width=0.49\columnwidth]{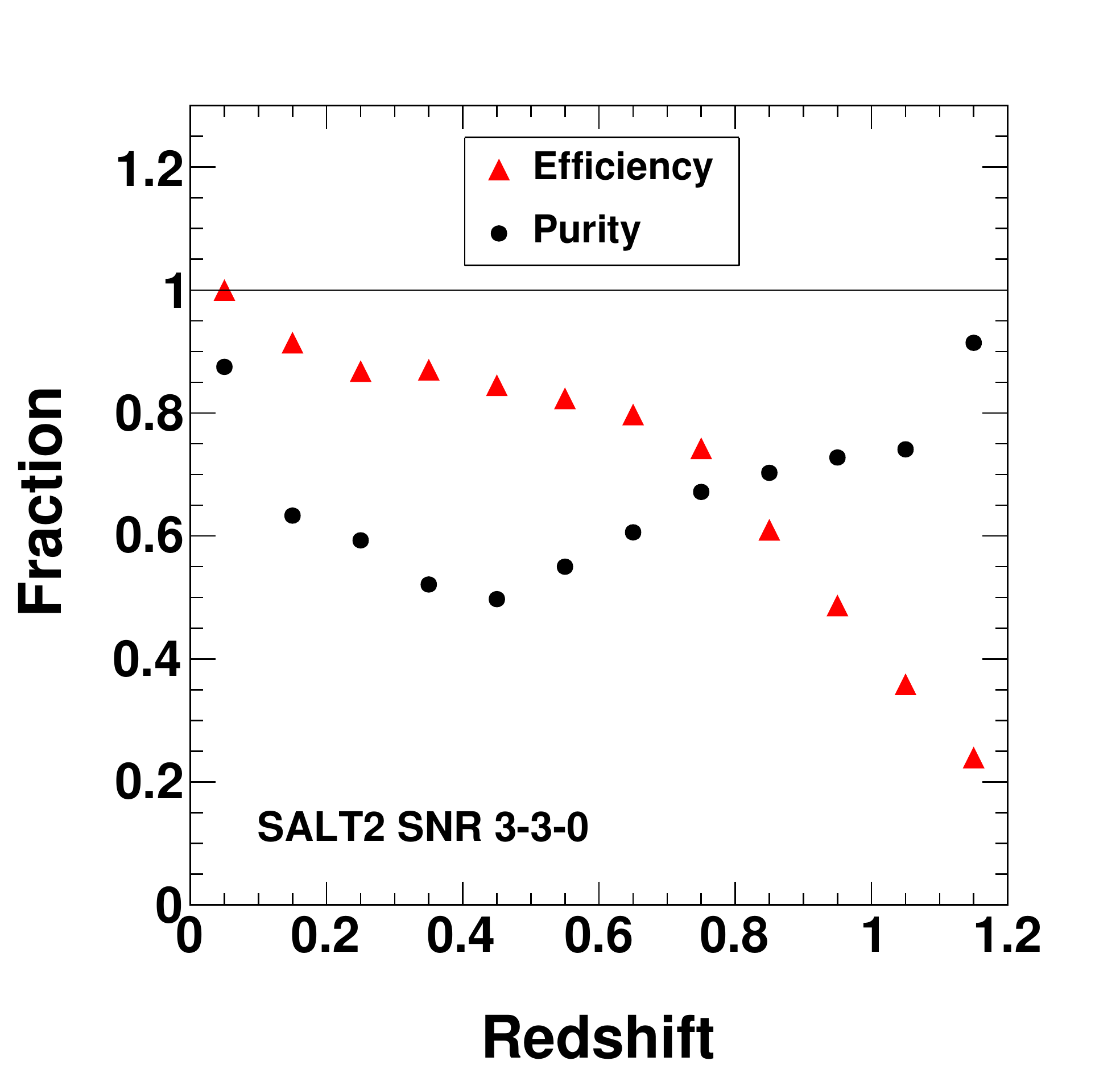}
\caption{The \mlcs\ purities and SNIa efficiencies are plotted in the left panels.
The definition of SNIa efficiency is the same as in Tab.~\ref{tab:purities}.
The top left panel is for the tightest SNR cuts, SNR-10-5-5,  while the bottom left panel is 
for the loosest set of SNR cuts, SNR-3-3-0.  
The corresponding \salt\ purities and SNIa efficiencies are shown in the right panels, 
with the tightest cuts on top and the loosest cuts on bottom.  The jumps in 
purity observed at some redshifts are due to low statistics in the SNcc passing
the cuts.}\label{fig:purities}
\end{center}
\end{figure}

In this section, we present the results for purities and 
SNIa efficiencies for the four sets of SNRMAX cuts  and with the 
\mlcs\ and \salt\ fit probability cuts described above.  
We define the SNIa efficiency as the ratio of the number of SNIa
passing all cuts that define the sample to the total number of SNIa simulated.
For our calculation of efficiency, the 
denominator is the complete sample of SNIa generated with zero SNR cuts
and the rates described in Ref.~\cite{bern11}.  
Many studies of SNIa efficiencies apply different sets of base 
SNR cuts, making it difficult to compare 
efficiencies from different analyses.  
We define the sample purity as the ratio of the number of 
SNIa to the total number of SNIa+SNcc passing all cuts.
The numbers of SNIa and SNcc and
the related purities and efficiencies integrated over all redshifts 
are presented in Tab.~\ref{tab:purities}. Figure~\ref{fig:purities} shows
the purities and efficiencies as functions of redshift for the 
tightest and loosest SNR cuts for both the \mlcs\ and \salt\ models.  
As discussed above
for Fig.~\ref{fig:fp_compare}, the \salt\ model without tight priors
is more flexible and leads to lower purities than our current implementation
of the \mlcs\ model. 

\begin{table}[ht]
\begin{center}
\small
\begin{tabular}{|c|c|c|c|c|c|} 
\hline
SNRMAX Cuts & Algorithm & SNIa & SNcc & Purity & Efficiency \\\hline \hline
SNR-10-5-5 & \fpmlcs $>0.1$ & 3534 & 88 & 98\% & 20\%\\
SNR-5-5-5 & \fpmlcs $>0.1$ & 4659 & 240 & 95\% & 27\%\\
SNR-5-5-0 & \fpmlcs $>0.1$ & 5949 & 534 & 92\% & 34\% \\
SNR-3-3-0 & \fpmlcs $>0.1$ & 9206 & 3138 & 75\% & 53\% \\
\hline
SNR-10-5-5 & \fpsalt $>0.1$ & 3686 & 236 & 94\% & 21\%\\
SNR-5-5-5 & \fpsalt $>0.1$ & 4820 & 568 & 89\% & 27\%\\
SNR-5-5-0 & \fpsalt $>0.1$ & 6425 & 1173 & 85\% & 37\% \\
SNR-3-3-0 & \fpsalt $>0.1$ & 9776 & 5298 & 65\% & 56\% \\
\hline
\end{tabular}
\caption{The simulated SNe sample purities and efficiencies are presented for a
variety of selection criteria (symbols defined in Tab.~\ref{tab:symbols})
and two SNIa identification methods (\fpmlcs\ and 
\fpsalt\ are the fit 
probabilities for the \mlcs\ and \salt\ models respectively).  
For the calculation of efficiency, the 
denominator is the complete sample of SNIa generated with 
no SNR cuts.
The number of input SNIa was 17555.}\label{tab:purities}
\end{center}
\end{table}

\begin{table}[h]
\begin{center}
\small
\begin{tabular}{|c|c|c|} 
\hline
Type & Bernstein {\it et al.} & This Analysis \\\hline \hline
Ib/c & 57 & 54 \\
IIP & 2 & 5 \\
IIn & 2 & 0 \\
IIL & 2 & 29 \\ 
\hline
\end{tabular}
\caption{The number of core collapse supernovae passing the 
tightest SNR-10-5-5 cuts and \fpmlcs $>0.1$, compared 
to Ref.~\cite{bern11}.  The large increase in the number of
Type IIL SNe passing cuts is mostly due to the 
0.5 magnitude brighter input value to the \snana\ template, coming
from Ref.~\cite{Li11}.
}\label{tab:types}
\end{center}
\end{table}

\begin{figure}[th]
\begin{center}
\includegraphics[width=0.49\columnwidth]{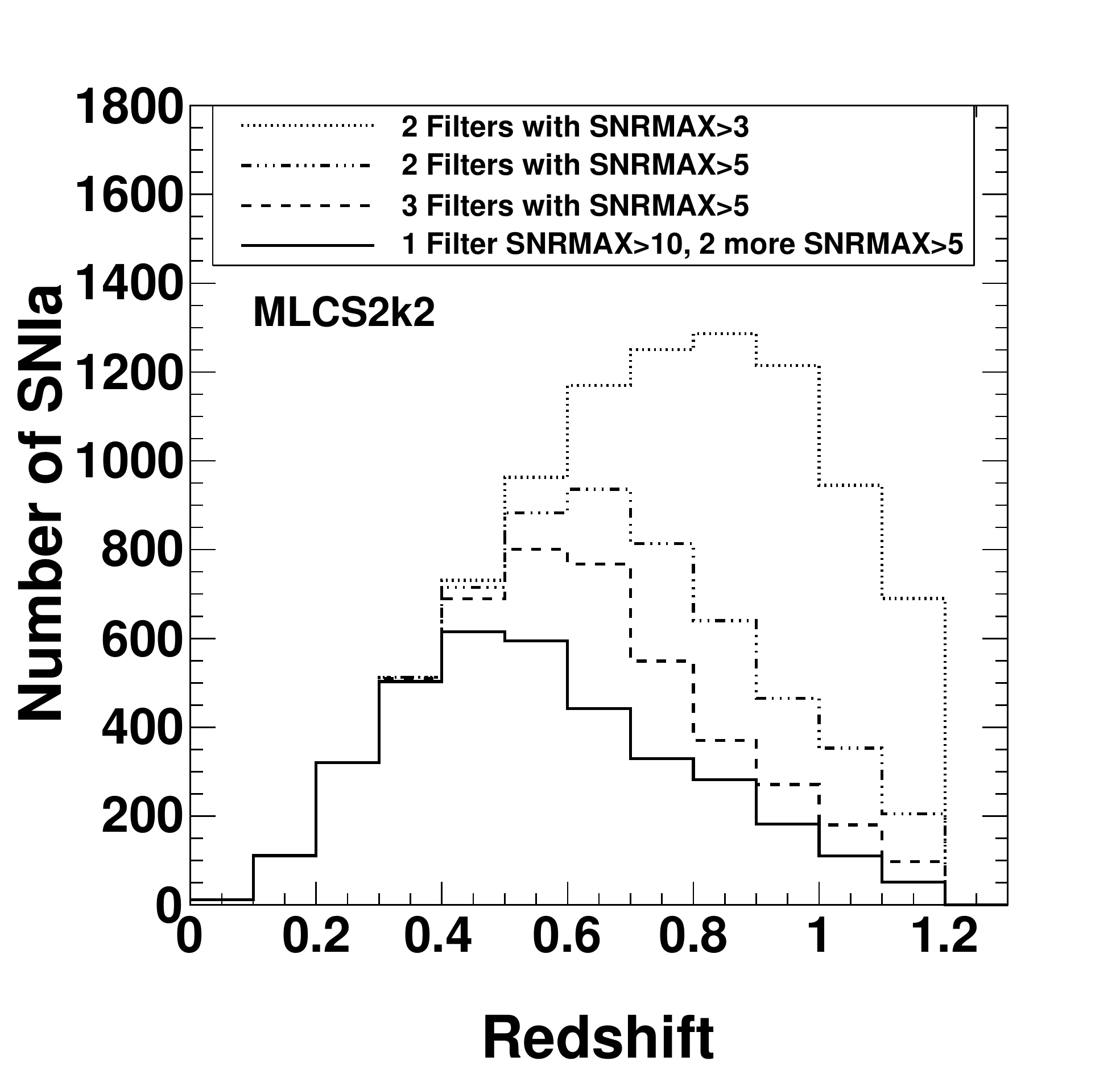}
\includegraphics[width=0.49\columnwidth]{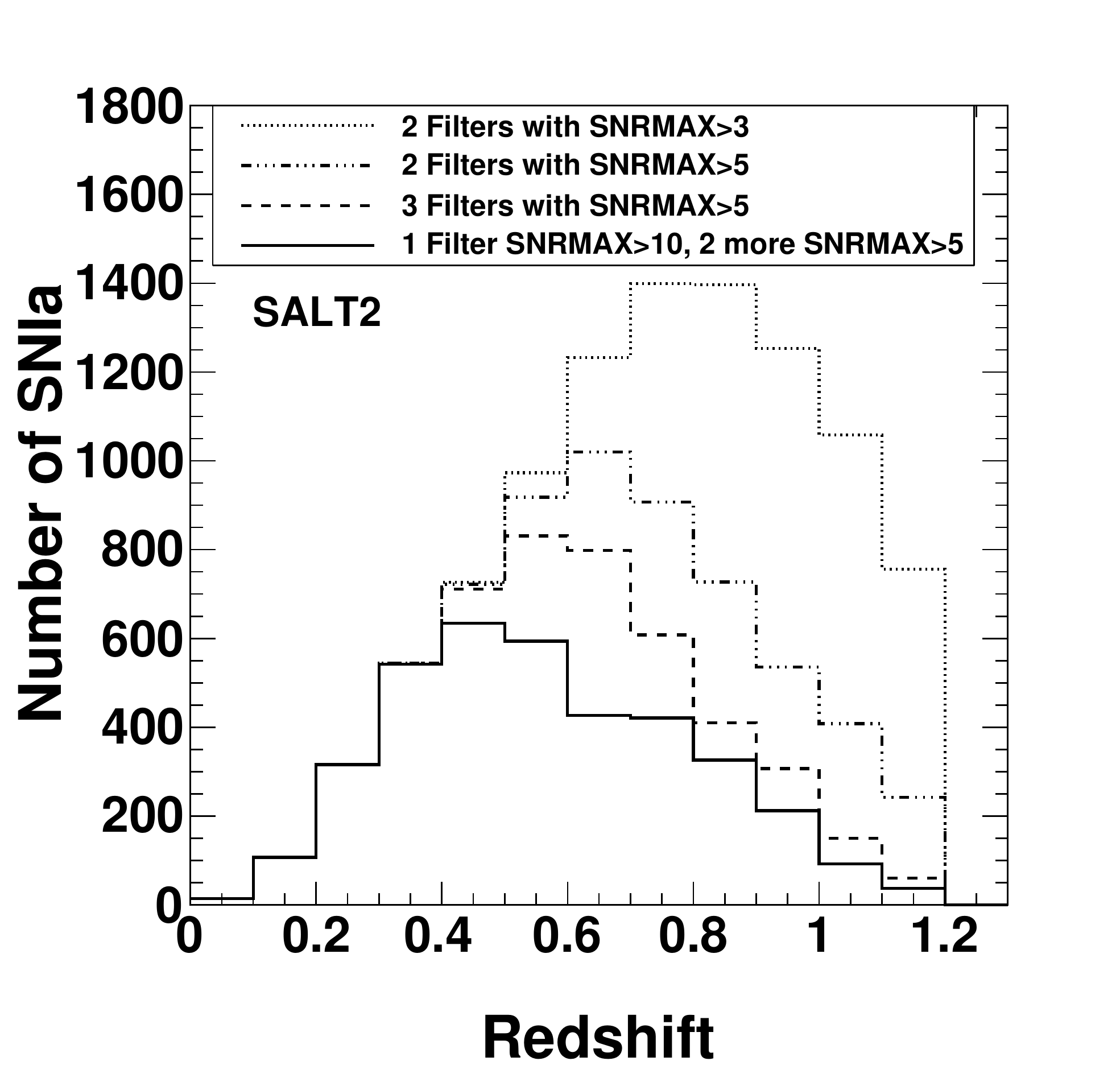}
\includegraphics[width=0.49\columnwidth]{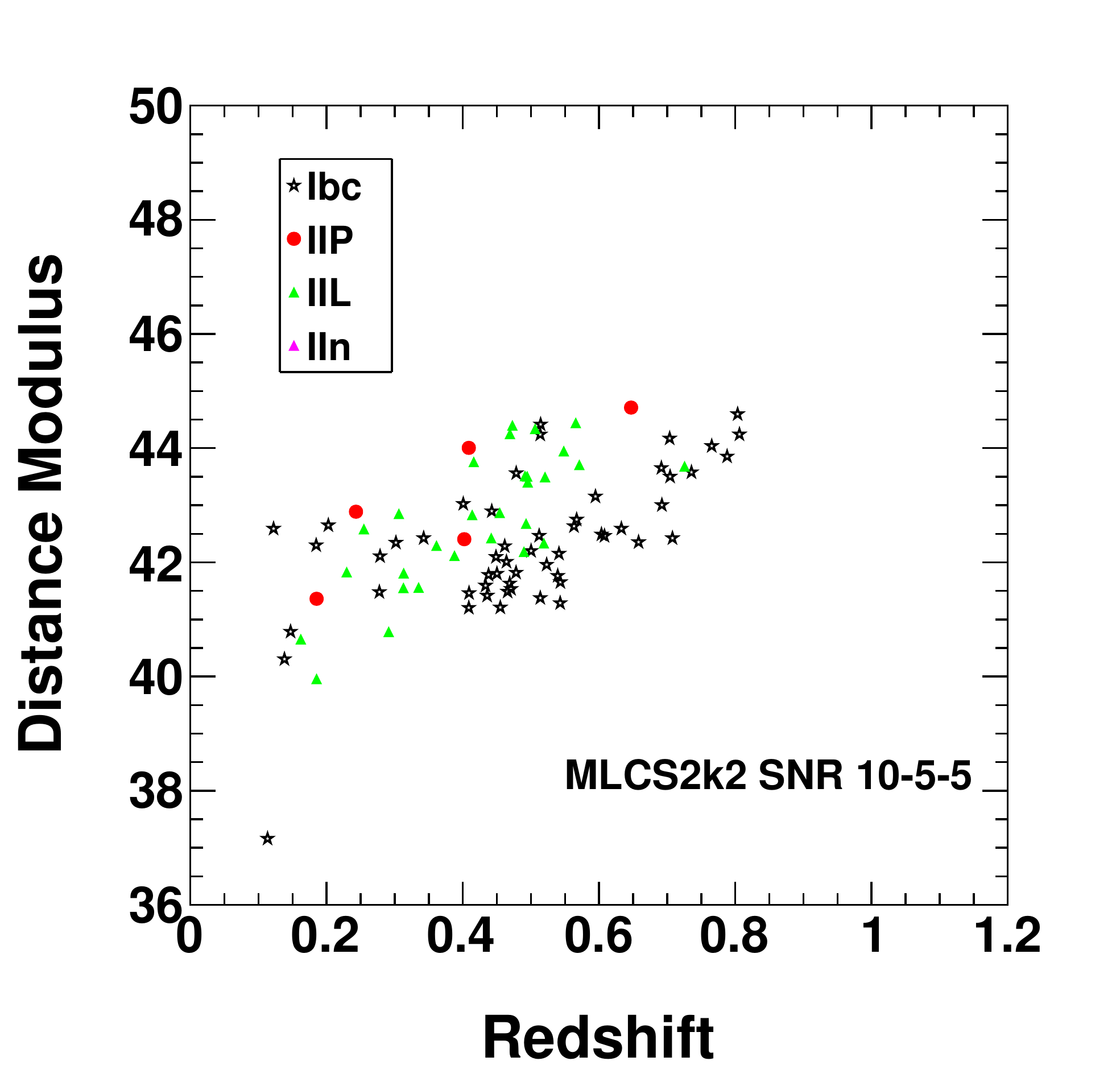}
\includegraphics[width=0.49\columnwidth]{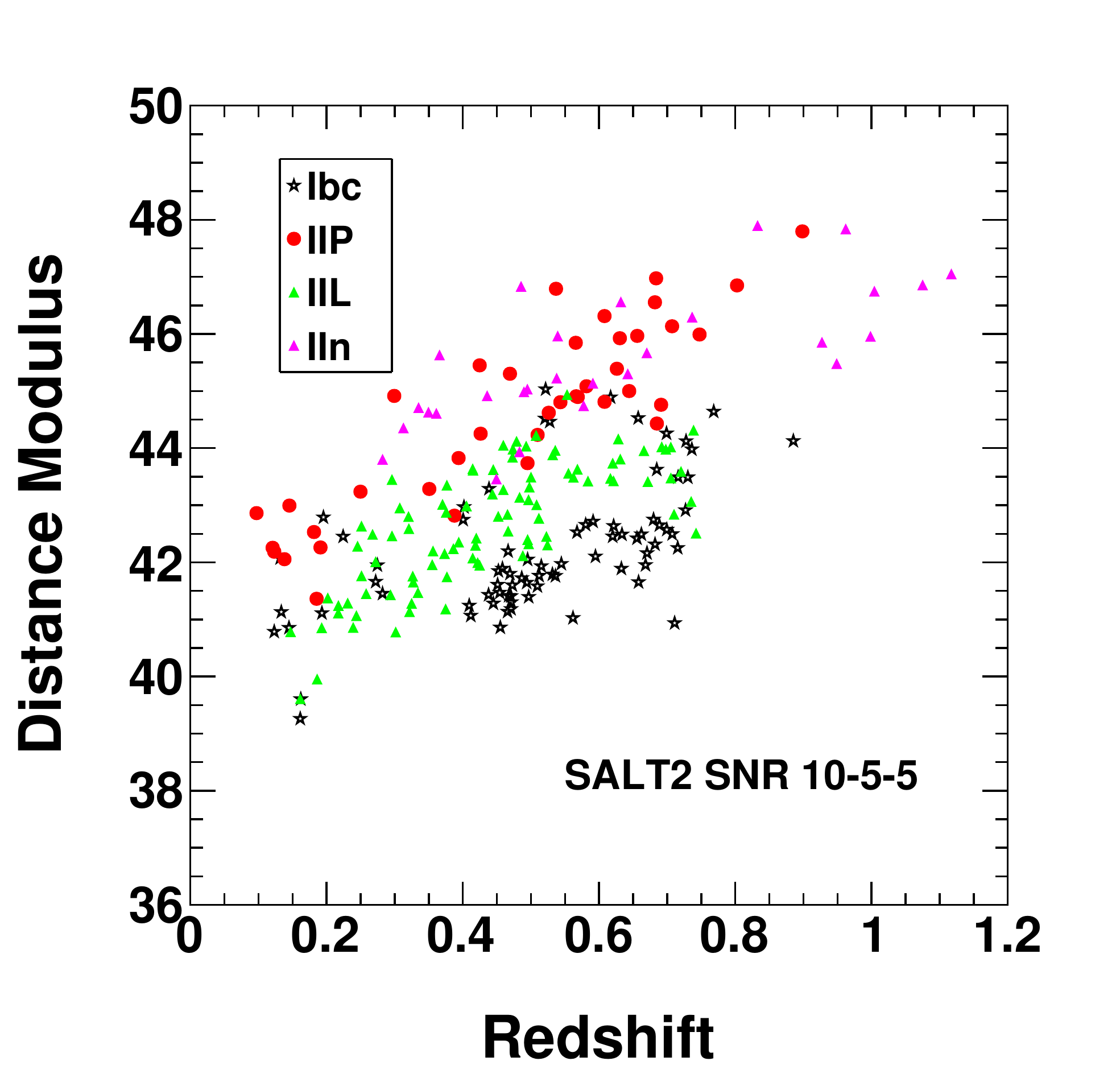}
\caption{The top panels show the SNIa redshift distributions for the \mlcs\ model (left) 
and the \salt\ model (right) for each SNRMAX cut and a fit probability$>0.1$ cut 
in each case.  The lower panels show the Hubble scatter for the SNcc sample passing
the tightest SNR-10-5-5 cut and a fit probability$>0.1$ cut in each case.
}\label{fig:zhists_and_types}
\end{center}
\end{figure}

Figure~\ref{fig:zhists_and_types} shows some characteristics of the SNIa and
SNcc samples that pass the fit probability cuts.   The SNIa redshift
distributions shown on the top panel demonstrate the increasing 
SNIa efficiency at large redshift as the cuts are relaxed.  The 
top left panel is for \mlcs\ and the top right panel is for the \salt\ model.
The Hubble scatter shown on the lower panels is for the SNcc sample passing
the SNR-10-5-5 cuts and a fit probability$>0.1$ cut in each case.
The number of type IIL SNcc is significantly 
more than in Ref.~\cite{bern11}.  The numbers for each SNcc type 
with the SNR-10-5-5 cuts are shown in Tab.~\ref{tab:types}. This change in 
SNcc numbers is due to the simulation input changes 
described in $\S$\ref{sec:siminput} and Appendix~A.  

\section{Dark Energy Task Force Figure of Merit}\label{sec:detffom}
\subsection{Figure of Merit Calculation}\label{sec:fomcalc}
As described in Ref.~\cite{bern11}, constraints on cosmological parameters are
obtained by comparing the predicted theoretical values of distance moduli, 
$\mu(z,\theta_c)$, to the values inferred from the light
curve fits of the SN simulations, $\mu_{obs}(z)$, where:
$\theta_{c} \equiv \{\Omega_{DE},w_0, w_a,\Omega_k\}$ is the set of 
cosmological parameters. Here, $\Omega_{DE}$ and $\Omega_k$ are the 
current energy densities corresponding to Dark Energy and spatial curvature as a 
fraction of the critical density. The parameters $w_0,w_a$ are the 
parameters in a CPL parametrization~\cite{che01,lin03} of the equation of state:
$w(a) = w_0 + w_a (1- a).$  
The likelihood for  an individual 
SN, at redshift z$_i$, is taken to be a
Gaussian with a mean given by the $\mu(z_i,\theta_c)$ at redshift z$_i$ 
for the cosmological parameters $\theta_c$. 
The simulated SN observations are independent and the likelihood is analytically 
marginalized over the nuisance-parameter combination of the Hubble 
Constant, $H_0$, and the absolute magnitude, $M$, with a flat prior on $M$. 
Following the DETF Report \cite{alb06}, 
we evaluated the 
performance of photometric identication algorithms and 
selection criteria in terms of the DETF FoM. 

The distance modulus errors used in the FoM calculation are
typically derived from the light curve fit errors and an intrinsic 
dispersion $\sigma_{int}=0.13$ added in quadrature as in Ref.~\cite{bern11}.
For a pure SNIa sample, the intrinsic dispersion is chosen to 
give $\chi^2/DOF\sim 1$ (DOF=number of degrees of freedom) in the cosmology 
fit. This procedure works for a very pure sample but will fail to 
account for the additional dispersion due to SNcc contamination.
Therefore, we will take an additional step in this
analysis of determining the RMS of the total SNIa+SNcc sample 
in each of 12 redshift bins. We then inflate the input errors for
the FoM calculation by adding an additional amount in quadrature to the reported errors
such that the means of the inflated errors match those RMS values. 
This ensures a reasonable $\chi^2/DOF\sim 1$ in the cosmology 
fit. Whenever we use inflated errors in the FoM calculation we
do not include the 0.13 intrinsic dispersion.

We model the issue of core collapse contamination in the Figure of Merit
calculation in a way similar to the method used in 
Ref.~\cite{bern11}. We assume that the distance modulus obtained by fitting a 
core collapse supernova to a Ia model is given by 
$$\mu_{cc}(\theta_c,z) = \mu(\theta_c,z) + \eta (z)$$
where $\mu(\theta_c,z)$ is the distance modulus and $\eta(z)$ 
encodes the differences in characteristics of core-collapse supernovae from 
SNIa and is independent of the cosmology. 
We expect $\eta(z)$ to be different for each core-collapse supernova 
 (certainly core collapse supernova types and templates), and hence  
there will be 
a large scatter in this quantity. For a DES simulation with a fixed 
set of selection cuts, we expect that $\langle\eta(z)\rangle$, the
 average value of $\eta(z)$ in each $z$ bin, to be roughly consistent 
between different realizations of simulations. We use our simulated data
to determine $\langle \eta(z) \rangle $  for each set of selection cuts 
used in this analysis. 

With this information from simulations, we obtain an average correction
 as a function of redshift for the average shift in $\mu_{obs}(z)$ 
 introduced by the  core-collapse contamination. We parametrize this correction
 by noting that for a mixed sample the average value of $\mu_{obs}(z)$ is 
given by  
\begin{eqnarray}
\langle \mu_{obs} \rangle &=& f(z)\langle\mu_{cc}\rangle + (1- f(z)) \langle\mu_{Ia}\rangle,\nonumber\\
   &=& \mu(\theta_c,z) + f_{cc} \left[F(z)\langle\eta(z)\rangle\right],
\label{eqn:muccequation}
\end{eqnarray}
where $f(z)$ is the probability that a randomly selected supernova at 
redshift $z$ in the mixed sample is a core collapse supernova,
$f_{cc} = 1 -$ purity, $F(z)=f(z)/f_{cc},$ and we have assumed that 
the observed distance moduli of SNIa are unbiased estimates of the 
distance moduli $\mu (\theta_c,z)$.  The average correction
 is the second term in Eqn.~\ref{eqn:muccequation} and the part in the square
brackets is computed from simulations and is held fixed. 
Thus, we expand our set of model parameters to include $f_{cc}$ and study 
joint constraints on all the parameters including $f_{cc}$. In order to obtain a 
Figure of Merit analogous to the DETF FoM, we evaluate a Fisher Matrix 
at the DETF fiducial values of the cosmological parameters and marginalize over
 all parameters other than $w_0$ and $w_a$. In doing so, we use priors 
on the model 
parameters. A Gaussian prior on $f_{cc}$ with a standard deviation 
assumed to be equal to $f_{cc}$ is used. The choices of priors on cosmological
parameters are the same as those used in Ref.~\cite{bern11}: 
a Fisher matrix representing priors on the set of cosmological parameters 
from DETF StageII experiments and expected Planck data was used 
along with a set of low redshift supernovae
from experiments that were not included 
in calculating the StageII Fisher matrix mentioned above. These low 
redshift SNe were spectroscopically identified,
and thus neither of the modifications for core-collapse supernovae (the 
inflation of the reported errors to match RMS or the use of the polynomial)
were applied to these supernovae.

\subsection{Factors affecting the Figure of Merit}
In this section, we investigate various factors that affect
the DETF Figure of Merit, and give a step-by-step example of how the FoM changes
with each factor.  We first investigate our sensitivity to 
the fit probability cut shown in Fig.~\ref{fig:fp_compare}.  
We show in Tab.~\ref{tab:fom2} how the FoM changes for
fit probabilities greater than 0.05, 0.1 and 0.2, 
for the tightest (SNR-10-5-5) and
loosest (SNR-3-3-0) cuts.
The FoM is relatively insensitive to the precise fit
probability cut for both \mlcs\ and \salt.  Therefore for the
rest of this paper, we will use 0.1 as the cut point, the
same as in Ref.~\cite{bern11}.

\begin{table}[ht]
\begin{center}
\begin{tabular}{|c|c|c|} 
\hline
SNRMAX cuts & ID algorithm & FoM SNIa$+$SNcc+Sys.\\\hline
\hline
SNR-10-5-5 & \fpmlcs $>0.05$  & 189 \\
SNR-10-5-5 & \fpmlcs $>0.1$   & 196 \\
SNR-10-5-5 & \fpmlcs $>0.2$   & 198 \\ \hline
SNR-3-3-0 &  \fpmlcs $>0.05$  & 159 \\
SNR-3-3-0 &  \fpmlcs $>0.1$   & 158 \\
SNR-3-3-0 &  \fpmlcs $>0.2$   & 155 \\ \hline
SNR-10-5-5 & \fpsalt $>0.05$  & 132 \\
SNR-10-5-5 & \fpsalt $>0.1$   & 132 \\
SNR-10-5-5 & \fpsalt $>0.2$   & 140 \\ \hline
SNR-3-3-0 &  \fpsalt $>0.05$  & 105 \\
SNR-3-3-0 &  \fpsalt $>0.1$   & 104 \\
SNR-3-3-0 &  \fpsalt $>0.2$   & 103 \\
\hline
\end{tabular}
\caption{The DETF Figure of Merit, including
SNcc systematics, is presented for a
variety of \mlcs\ and \salt\ fit probability selection criteria 
and for the loosest and tightest cuts considered \label{tab:fom2} and 
found to be relatively insensitive to exact value of the fit probability 
thresholds.}
\end{center}
\end{table}

\begin{figure}[h]
\begin{center}
\includegraphics[width=0.49\columnwidth]{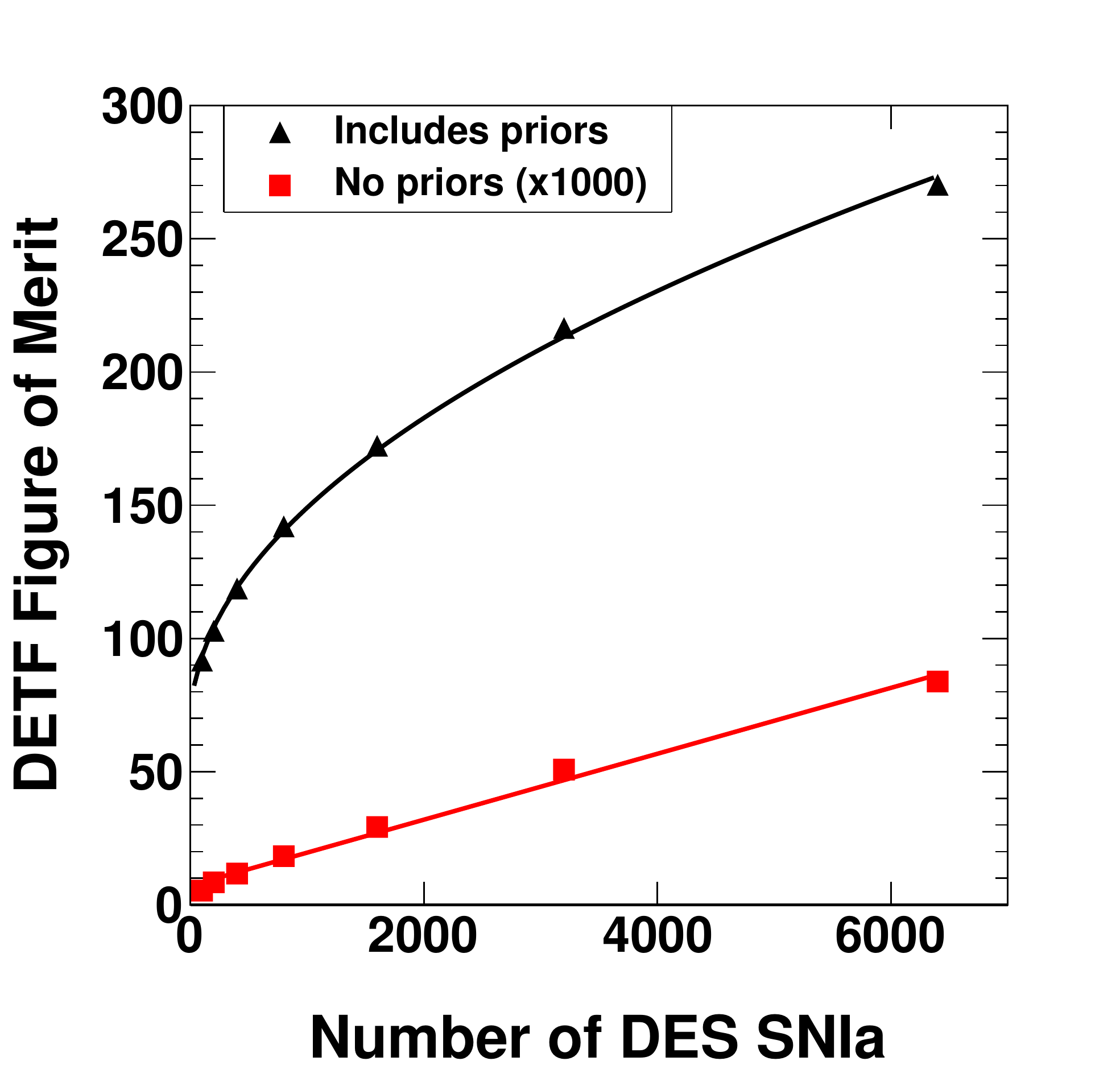}
\includegraphics[width=0.49\columnwidth]{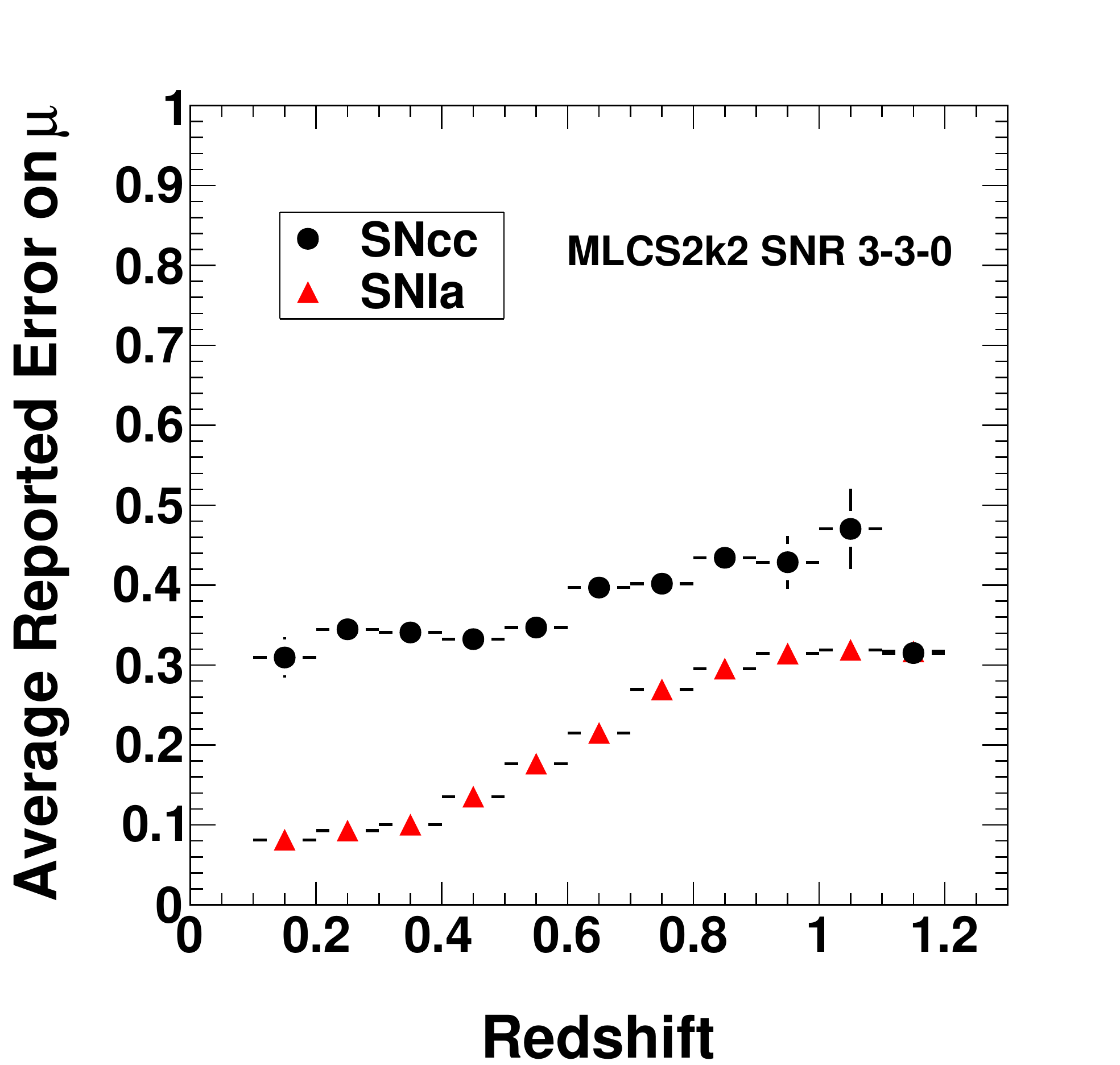}
\caption{The Figure of Merit dependence on the number of Type Ia SNe with and without
DETF Stage II and Planck priors is shown in the left panel. Note that the 
points without priors are multiplied by 1000.
The average $\mu$ error, as reported by the \mlcs\ Type Ia model fit, is shown 
for
the SNIa and SNcc samples in the right panel. Note that the highest redshift point
in the right panel has fluctuated so that the SNcc and SNIa points 
overlap and the SNIa point is not visible.}
\label{fig:NSNandmuerrors}
\end{center}
\end{figure}

We now investigate four factors that significantly affect the 
DETF Figure of Merit:
\begin{itemize}
\item the number of SNIa ($N_{Ia}$),
\item the reported errors on the distance modulus,
\item the effect of the additional Hubble scatter due to SNcc, which is 
usually much larger than the reported errors on the distance modulus,
\item the uncertainty in the Hubble residual due to SNcc systematics.
\end{itemize}
The number of SNe in the sample is an obvious factor
to consider. First, let us consider the simple DETF FoM without any of
our modifications. Here, adding supernovae to a sample will always improve
the FoM. If the supernovae added have the same statistical properties (similar
 error bars) then the rate of improvement with numbers depends on the priors
used in the calculation. This dependence can be observed in the left 
panel of Fig.~\ref{fig:NSNandmuerrors}, where independent statistically 
equivalent samples of SNIa passing selection cuts SNR-10-5-5 
(such as would be obtained by using more seasons of DES observation using 
the same selection cuts) have been added.
The lower curve, approximately linear\footnote{The linear dependence on
${N_{Ia}}$ can be understood from the fact that the DETF FoM is 
inversely proportional to the product of errors 
on $w_0$ and $w_a$, each of which fall like $\sqrt{N_{Ia}}$.},
 is the variation of the 
FoM with no DETF Stage II or Planck priors,  and it has been multiplied
by 1000 in order to be visible.  That demonstrates how critically important
the priors are in the FoM.  The upper curve includes the DETF stage II and 
Planck priors and has a $\sqrt{N_{Ia}}$ dependence\footnote{This dependence 
is difficult to predict due to the complicated effects of the priors.}.

We begin our step-by-step example of FoM changes with the top row of 
Tab.~\ref{tab:example}, for the loosest cuts SNR-3-3-0 and 
\fpmlcs\ $>0.1$. While 9206 SNIa pass these cuts, these SNIa are not 
statistically equivalent to the SNIa used in the left panel 
of Fig.~\ref{fig:NSNandmuerrors}, as supernovae
that fail the tightest cuts have larger errors on the distance modulus.  
The FoM value (with priors) of 310 is for the 9206 SNIa that pass
these cuts with the reported distance moduli errors added in quadrature
with 0.13,  as discussed above, and used in Ref.~\cite{bern11}.
The value extrapolated from the left panel of Fig.~\ref{fig:NSNandmuerrors} 
is 316. The difference is even starker for second row of the table, where the FoM=354 is the 
extrapolated value from the left panel when the number of SN is raised to 
12344, the total number of supernovae (SNIa and SNcc) in the sample. 
The third row of the table shows
the calculated FoM of 316 using 0.13 added in quadrature to the reported distance
moduli errors.  Most of the difference between 354 and 316 is due to the reported errors on the distance 
moduli of SNcc being larger than SNIa, as shown in the right panel of 
Fig.~\ref{fig:NSNandmuerrors}.

\begin{figure}[th]
\begin{center}
\includegraphics[width=0.49\columnwidth]{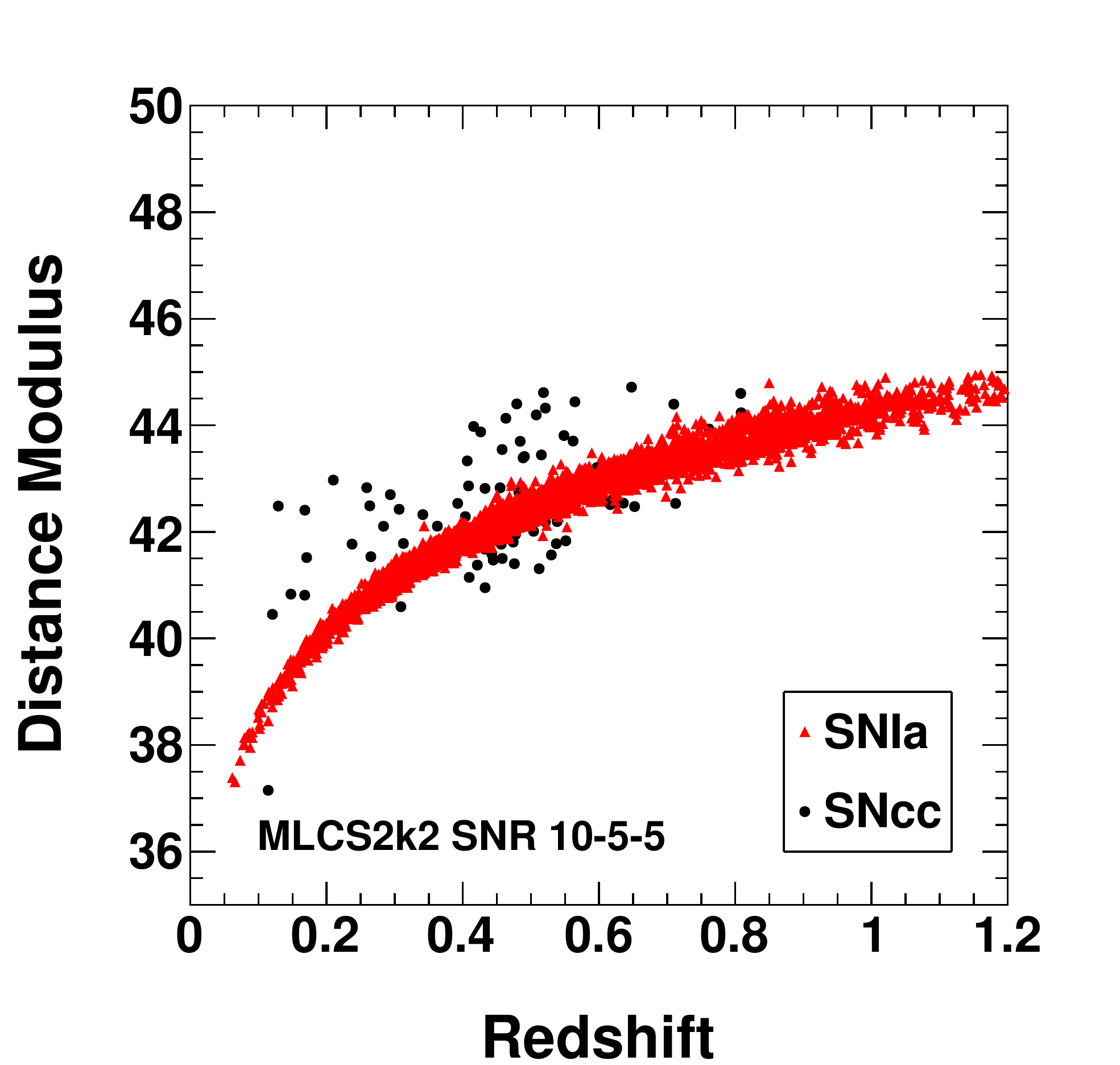}
\includegraphics[width=0.49\columnwidth]{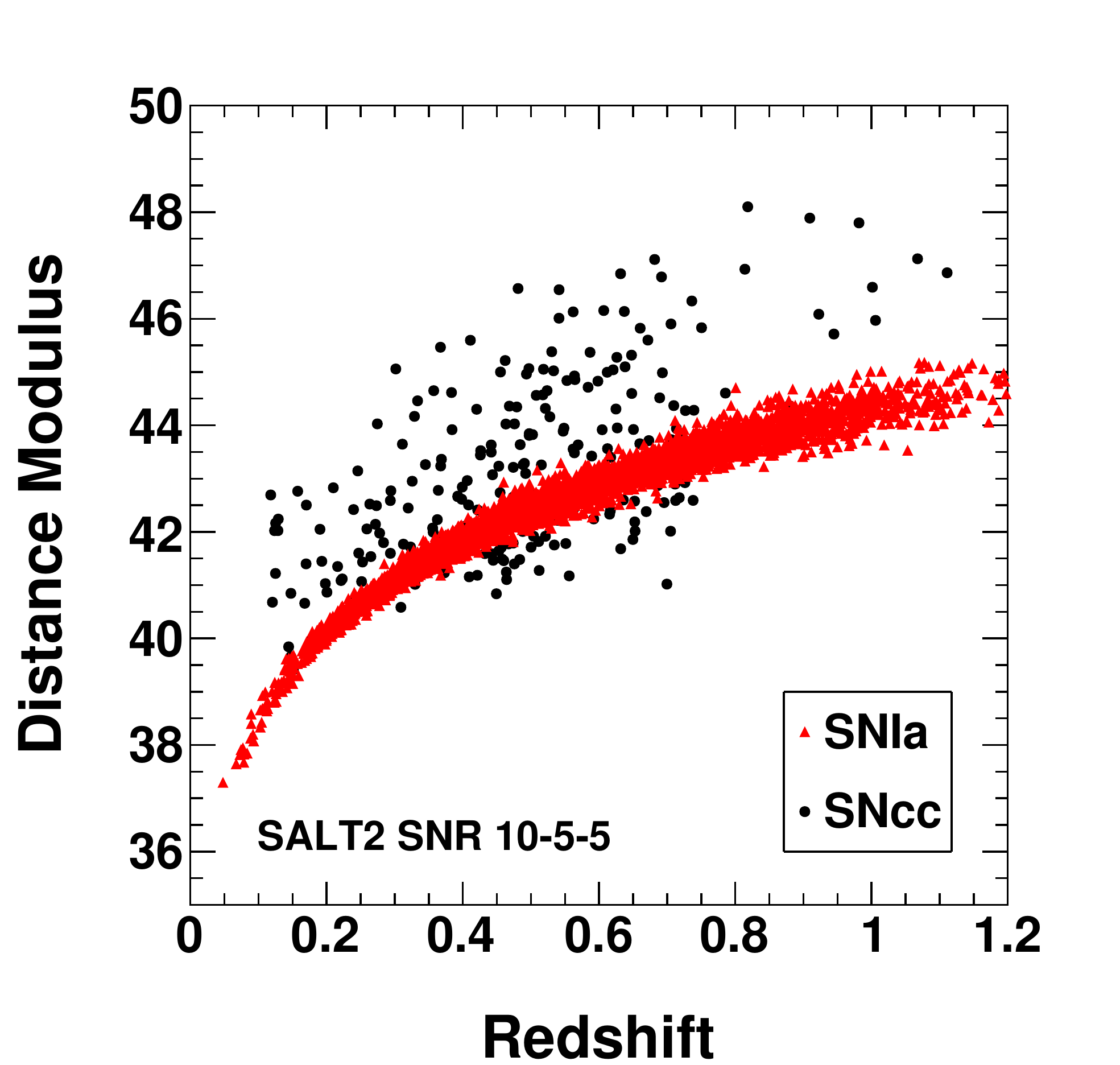}
\includegraphics[width=0.49\columnwidth]{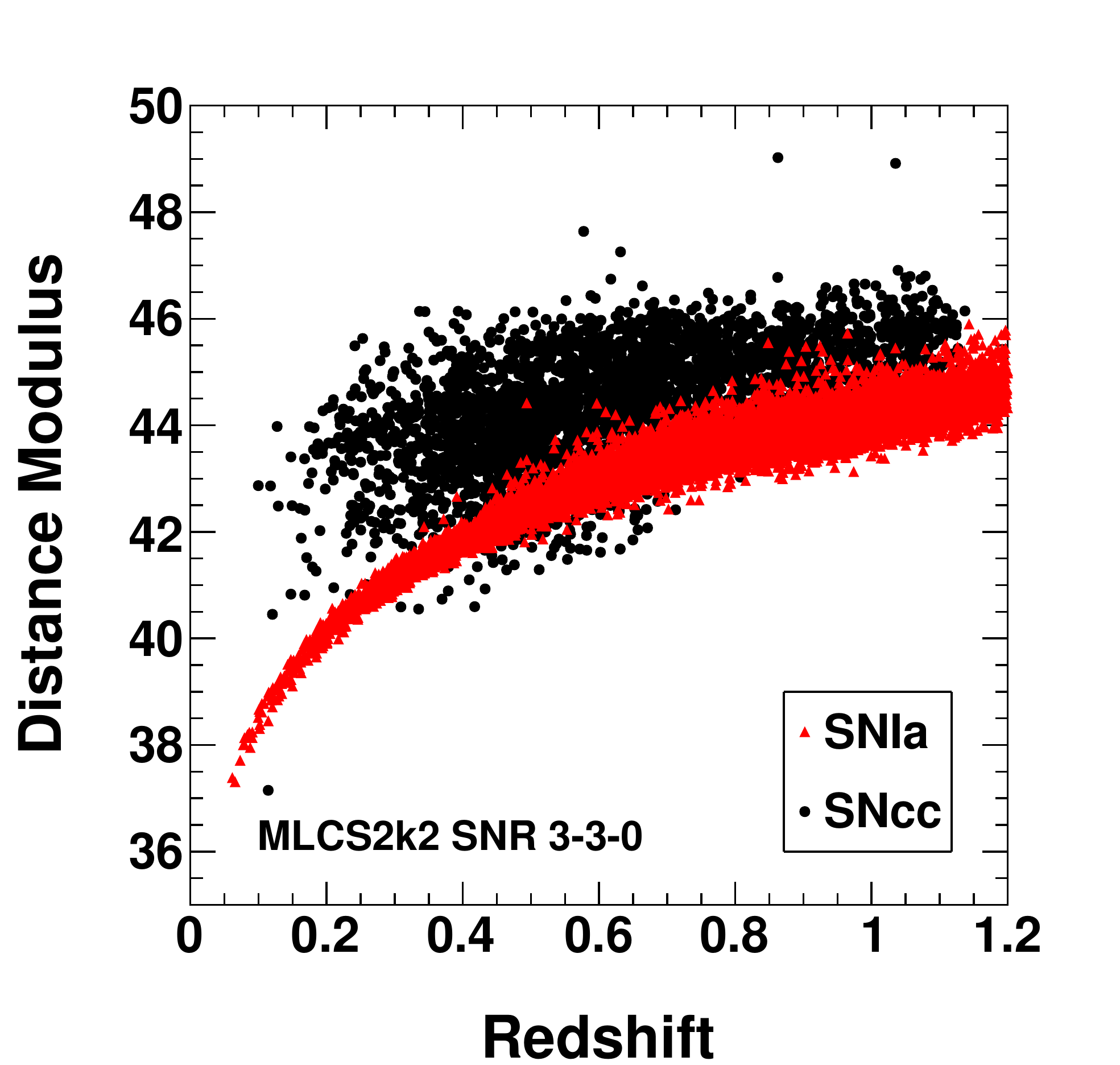}
\includegraphics[width=0.49\columnwidth]{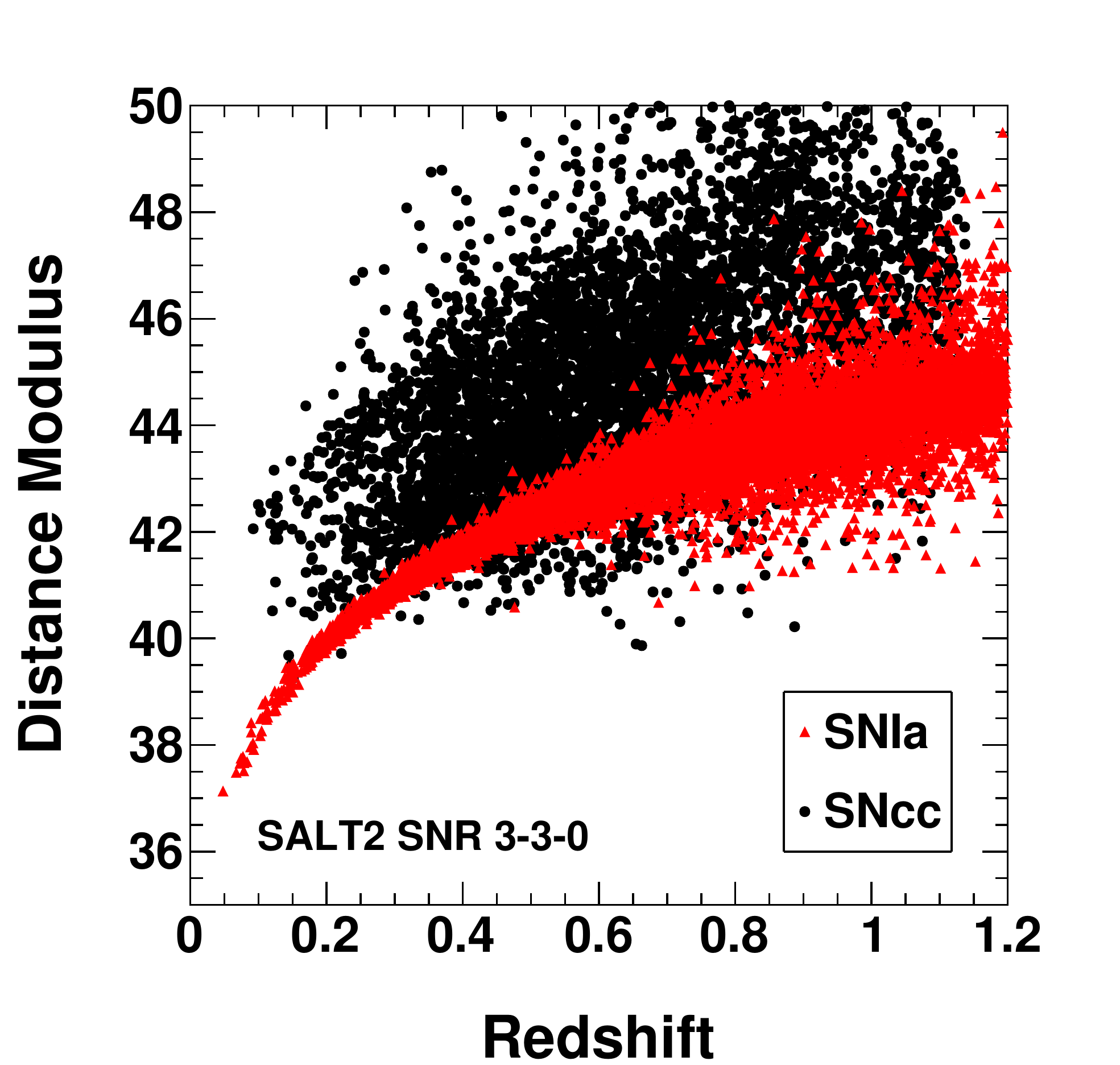}
\caption{We display four Hubble diagrams, for both SNIa and SNcc. The top panels are 
for the tightest SNR-10-5-5 cuts,  while the lower panels are for the loosest SNR-3-3-0 
cuts.  The left panels are with the \mlcs\ light-curve fitter, while the right panels
are for the \salt\ light-curve fitter. (See text for discussion.) }
\label{fig:hubblescatter}
\end{center}
\end{figure}

The third factor affecting the FoM is the effect of the SNcc
scatter on the Hubble diagram, demonstrated in Fig.~\ref{fig:hubblescatter}
for the loosest and tightest SNR cuts for both \mlcs\ and \salt\ models.
As discussed earlier,  we inflate the SNIa and SNcc 
input errors to the FoM determination in order to 
achieve a reasonable $\chi^2/DOF$.  The additional error added, for the 
loosest SNR-3-3-0 cuts, varies from 
0.8-1.5 mags for \salt\ and 0.3-1.3 mags for the \mlcs\ model.   Our choice of the loosest SNR cuts and 
\mlcs\ for our example shows the most dramatic effect of these
additional errors in the fourth row of Tab.~\ref{tab:example}.  This
 shows the FoM decreasing from 316 to 181.  
Note that this is just an exercise to understand 
the FoM better.  We do not expect anyone to attempt to do a cosmology 
analysis with the SNcc contamination shown in Fig.~\ref{fig:hubblescatter}
with the loosest cuts.

\begin{figure}[h]
\begin{center}
\includegraphics[width=0.49\columnwidth]{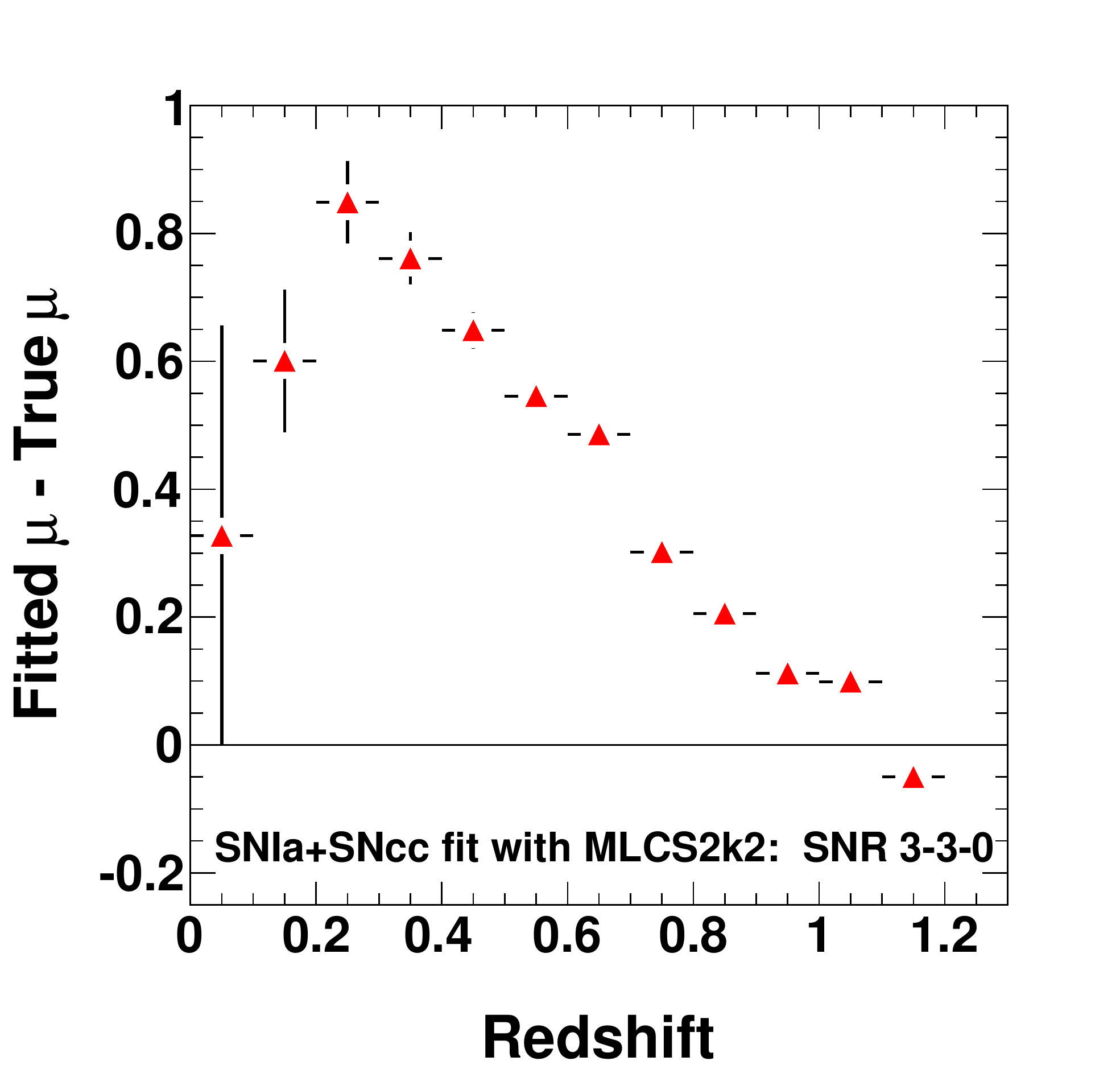}
\includegraphics[width=0.49\columnwidth]{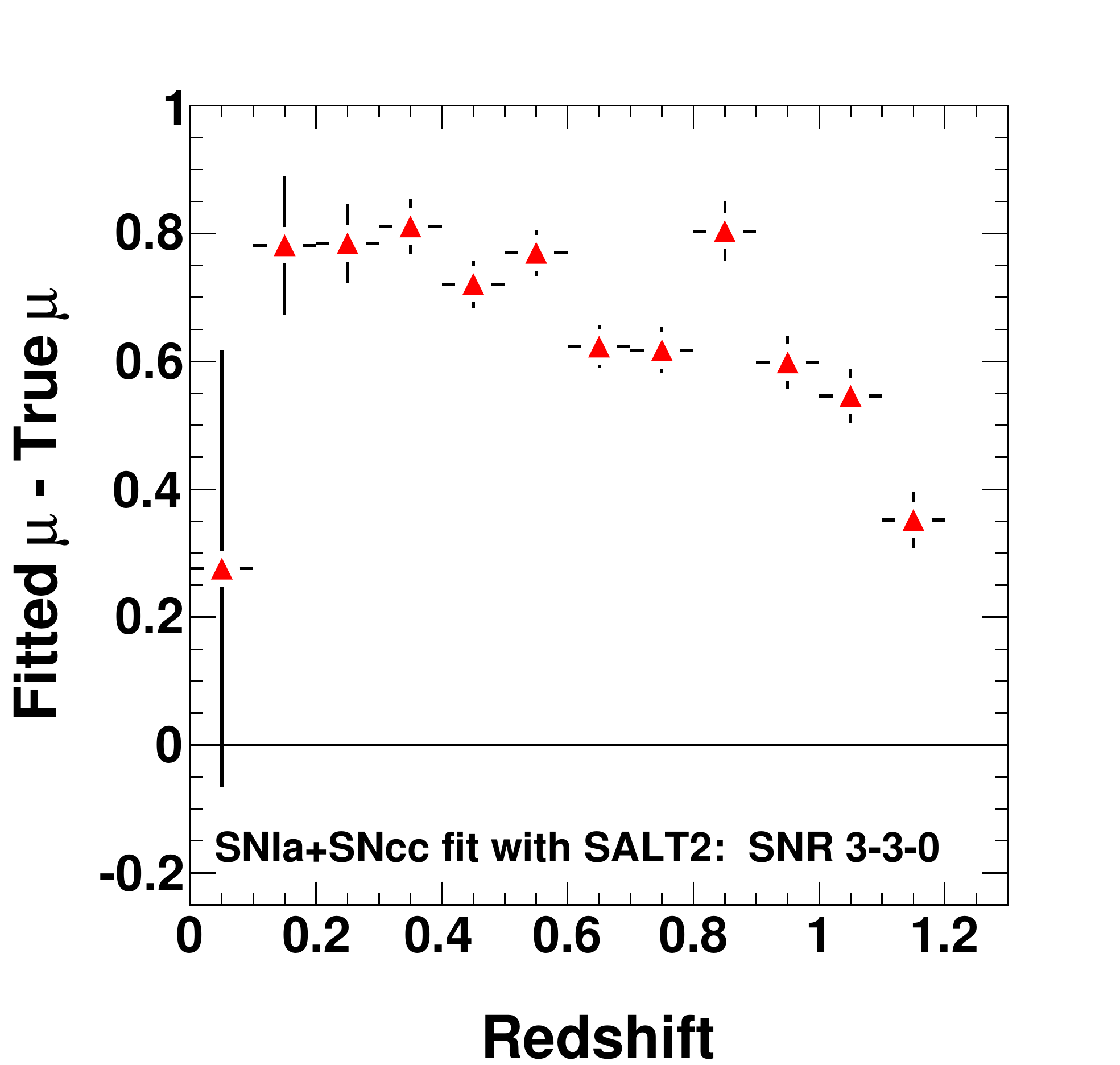}
\caption{The Hubble residual is shown when the core collapse sample is included, for the 
\mlcs\ Type Ia model on the left and the \salt\ model on the right, both with the loosest SNR-3-3-0 cuts.}
\label{fig:CCsystematic}
\end{center}
\end{figure}

The fourth and final SNcc-related factor we consider that affects the FoM
significantly is the strength of SNcc contamination. 
As discussed above and in Ref.~\cite{bern11}, 
we model as a function of $z$, the change of the Hubble residual 
due to varying amounts of SNcc contamination, and take 100\% of 
this change as the one standard deviation uncertainty. 
Figure~\ref{fig:CCsystematic} shows the
change in Hubble residual due to the SNcc sample for the loosest cuts and 
for the \mlcs\ and \salt\ models. 
The bottom row of Tab.~\ref{tab:example} shows the FoM decreasing 
from 181 to 158 using the change in Hubble residual with \mlcs\ 
and the loosest cuts.

\begin{table}[h]
\begin{center}
\small
\begin{tabular}{|l|c|}\hline
FoM for Ia only & 310 \\
\hline
FoM for SNIa + SNcc (Extrapolating using $\sqrt{N_{Ia}}$ dependence) & 354 \\
\hline
FoM for SNIa + SNcc(using fitter reported SNe $\mu$ error + 0.13 intrinsic dispersion) & 316 \\
\hline
FoM for SNIa + SNcc(inflated error for $\chi^2/DOF\sim 1$) & 181 \\
\hline
FoM for SNIa + SNcc(inflated error for $\chi^2/DOF\sim 1$ and SNcc systematic) & 158 \\
\hline
\end{tabular}
\caption{The most important factors that can change the DETF Figure of Merit 
are demonstrated with an example for \mlcs.  
(See the text for a detailed discussion.)}\label{tab:example}
\end{center}
\end{table}

Table~\ref{tab:fom1} shows a summary of our results, combining the four factors that
affect the FoM discussed above. We present different values for the FoM, 
from tighter cuts (SNR-10-5-5) to looser cuts (SNR-3-3-0), for both \mlcs\ and \salt\ models.
The column labeled FoM Ia is for SNIa and incorporates the reported SNIa 
$\mu$ errors added to 0.13 in quadrature.  The column labeled FoM Ia+CC 
adds the SNcc sample and includes the inflated error to make
$\chi^2/DOF\sim 1$ for each sample. The final column adds the SNcc systematic
described above and is the most relevant column.  The FoM is 
best for the tightest cuts for both \salt\ and \mlcs\ models. 
SNR cuts tighter than SNR-10-5-5 were also analyzed.
It is expected that as the SNR cuts are tightened, and purities
approach 100\%, the loss in SNIa statistics will cause a drop in FoM.
We observe this for both \mlcs\ and \salt, but at different SNR cut
values in each case.  However, for both \mlcs\ and \salt, the
peak FoM value occured at 98\% purity (SNR-10-10-5 and 
SNR-10-5-5 had almost identical purity and FoM for \mlcs\, 
and SNR-10-10-10 had the peak FoM for \salt).

We observe again that the \salt\ fitter allows more SNcc
into our sample than \mlcs\ and results in lower purities 
for a given set of SNR cuts. 
In addition, the Hubble scatter for the SNcc is larger
for \salt\ than \mlcs.  The lower
purity and larger scatter cause lower FoM for the 
samples fit with the \salt\ model.  As mentioned earlier, this is at least
partly due to the tight dust extinction priors 
used in the \mlcs\ fits~\cite{bern11}.

\begin{table}[h]
\begin{center}
\small
\begin{tabular}{|c|c|c|c|c|c|c|} 
\hline
SNRMAX cuts & SNIa ID & SNIa & Pur. & FoM Ia & FoM Ia$+$CC & FoM Ia$+$CC+Sys.\\\hline \hline
SNR-10-5-5 & \fpmlcs $>0.1$ & 3534 & 98\% & 249 & 203 & 196 \\
SNR-5-5-5  & \fpmlcs $>0.1$ & 4659 & 95\% & 266 & 200 & 172 \\
SNR-5-5-0  & \fpmlcs $>0.1$ & 5949 & 92\% & 285 & 197 & 167 \\
SNR-3-3-0  & \fpmlcs $>0.1$ & 9206 & 75\% & 310 & 181 & 158 \\
\hline
SNR-10-5-5 & \fpsalt $>0.1$ & 3686 & 94\% & 234 & 147 & 132 \\
SNR-5-5-5  & \fpsalt $>0.1$ & 4820 & 89\% & 246 & 151 & 131 \\
SNR-5-5-0  & \fpsalt $>0.1$ & 6425 & 85\% & 263 & 144 & 120 \\
SNR-3-3-0  & \fpsalt $>0.1$ & 9776 & 65\% & 276 & 130 & 104 \\
\hline
\end{tabular}
\caption{The Dark Energy Task Force Figure of Merit is presented for a
variety of selection criteria (symbols defined in Tab.~\ref{tab:symbols})  
and fits to two SNIa models. The columns, from left to right, are: 
cut value, SNIa model used, number of SNIa passing all cuts, 
purity of the sample, the FoM for a SNIa-only sample and statistical 
uncertainties only, the FoM for the combined SNIa+SNcc sample
using the inflated errors for the complete sample,  and the FoM 
for the combined sample including the core collapse 
systematic uncertainty.
SNRMAX cuts tighter than those shown in this table were 
investigated.  For both \mlcs\ and \salt, the FoM decreased 
for purities higher than 98\%.
}\label{tab:fom1}
\end{center}
\end{table}


\section{Discussion}\label{sec:discuss}
This analysis is an extension of the Bernstein {\it et al.}~\cite{bern11} 
paper with a focus on the Type Ia/core collapse separation.
The previous analysis used one set of SNR cuts and one SNIa 
identification model and found a negligible effect on 
cosmology. On the other hand, the tight cuts led to a low 
efficiency for SNIa.  It is natural to investigate 
additional SNR cuts and models.

The main extensions to Ref.~\cite{bern11} addressed in this analysis 
are listed below:

\begin{itemize}
\item four sets of SNR cuts are used instead of one,
\item the \mlcs\ and \salt\ models are treated on equal footing,
\item the \snana\ simulation inputs for core collapse simulations 
have been updated with more current knowledge (a total of seven input changes 
were made and are detailed in Appendix~A),
\item a variety of fit probability and purity/efficiency plots are presented,
\item a detailed purity and efficiency table for all variations is presented,
\item the Dark Energy Task Force Figure of Merit is tested on all variations,  
\item the normal procedure of adding an intrinsic dispersion $\sim$ 0.13 in quadrature
with the  reported $\mu$ error is supplemented  by a new procedure that uses 
inflated errors determined from the Hubble diagram RMS of the total SNIa+SNcc sample in redshift bins,
\item other significant factors in the FoM are examined, such as the scaling 
with number of supernovae and large impact of the DETF stage II and Planck priors (more than x1000 
increase in FoM), as well as the core collapse systematic uncertainty.
\end{itemize}

The changes to the simulation inputs caused very little change in 
resulting purities, compared to Ref.~\cite{bern11}.  The most 
significant change is in the Type IIL simulation,  which increased
in number passing cuts by almost a factor of 15.  This is mostly due to the 
0.5 magnitude brighter input value to the \snana\ template, coming
from Ref.~\cite{Li11}.  This highlights a fragility in the 
current knowledge of core collapse simulations, there is 
only one Type IIL template available to generate the simulations.  
Hopefully this can be supplemented by more templates in the future.

We find the \salt\ model allows more SNcc passing the fit probability cuts
compared to the \mlcs\ model (which includes dust extinction priors) 
and leads to somewhat lower purities 
and lower FoM. The scatter in the SNcc $\mu$ values is also larger
for the \salt\ model, and this also contributes to lower FoM 
with our treatment of core collapse uncertainties.  
The more significant result, however, is that for both models the FoM decreases 
with purities lower than 98\%.
Purities higher than 98\% were analyzed and the FoM 
decreased for both models due to a loss of SNIa statistics.

This analysis lays the groundwork for future analyses of more sophisticated 
photometric typers~\cite{SNchall}, as well as the application of additional 
cuts, such as SNe color and stretch.  From the top panels in 
Fig.~\ref{fig:zhists_and_types}, it is clear that the SNIa sample 
is complete up to $z \sim 0.4$ even with the tightest SNR cuts.
Therefore, loosening SNR cuts for low redshift only increases
the SNcc contamination. Furthermore, at $z\gtrsim 0.8$ the purity 
is increasing even with the loosest cuts.
These trends imply that using a $z$-dependent SNR cut would
be a better choice.

In addition,  it is obvious from the Hubble scatter plots that 
many of the SNcc are easily removed, as their $\mu$ values are many standard deviations 
away from any possible cosmology.  Coupled with potential $\mu$ cuts, it is 
necessary to study fitted cosmology biases for each possible sample, since the DETF FoM is not
sensitive to these biases.  Another challenging future study is the 
measurement of SNe colors (either \salt\ $\beta$ or \mlcs\ $R_V$), as a function of redshift, in the 
presence of varying amounts of SNcc contamination.
Finally, the treatment of the core collapse 
systematic (100\% of the shift in Hubble residual) is simplistic and 
can be improved with a breakdown of the individual components of the
core collapse uncertainties.

\acknowledgments

We thank Rick Kessler for his advice concerning the \snana\ simulations used in this paper.
We thank Kyler Kuehn for a critical reading of the manuscript.

The submitted manuscript has been created by UChicago Argonne, LLC, 
Operator of Argonne National Laboratory (``Argonne''). Argonne, a 
U.S. Department of Energy Office of Science laboratory, is operated under 
Contract No. DE-AC02-06CH11357. The U.S. Government retains for itself, 
and others acting on its behalf, a paid-up nonexclusive, irrevocable 
worldwide license in said article to reproduce, prepare derivative 
works, distribute copies to the public, and perform publicly and 
display publicly, by or on behalf of the Government.

\appendix
\section{Simulation Input Details}\label{apdx:siminputs}
In $\S$\ref{sec:siminput} we listed seven improvements to the core collapse
simulations, implemented since Ref.~\cite{bern11}. In this appendix, we 
provide more details for the experts in the field that are interested
in reproducing or expanding upon our simulations.

Measurements of the relative fractions of SNcc have improved recently, 
largely due to the LOSS data presented in Ref.~\cite{Li11}. Table~\ref{tab:fractions} 
compares the relative fractions of SNcc between Smartt {\it{et al.}}~\cite{sma09} and 
Li {\it{et al.}}~\cite{Li11}.  

\begin{table}[ht]
\begin{center}
 \small
\begin{tabular}{|c|c|c|} 
\hline
SNe Type & Smartt {\it et al.} \% & Li {\it et al.} \% \\\hline \hline
Ib & 9.8 & 5.2  \\
Ic & 19.6 & 13.3  \\
Ibc-pec & N/A & 6.0  \\
II-P & 58.7 & 52.7  \\
II-L & 2.7 & 7.3  \\
IIb & 5.4 & 9.0  \\
IIn & 3.8 & 6.5  \\
\hline
\end{tabular}
\caption{Comparison of core collapse relative fractions from 
Smartt {\it et al.}~\cite{sma09} and Li {\it et al.}~\cite{Li11}.}\label{tab:fractions}
\end{center}
\end{table}

We have made modifications in these relative SNcc fractions for our analysis since 
the Bernstein {\it{et al.}} paper:

\begin{itemize}
\item we use the Li {\it{et al.}} fractions with their inherent higher statistics 
and better sample completeness instead of the Smartt {\it{et al.}} fractions,
\item for the Type Ib and Ic fractions, we take half 
of the total Ibc-pec (6\%) fraction and add this percentage (3\%) to each of the Ib 
and Ic fractions (This is justified since Li {\it{et al.}} reported that the 
photometric behaviors of these SN Ibc-pec are all reasonably represented by the 
average SN Ibc light curve.),
\item we combine the II-L and IIb samples (7.3\% + 9.0\% = 16.3\%), since there are no IIb 
templates available currently in the \snana\ package.
\end{itemize}

The input relative fractions to \snana\ are summarized in Table~\ref{tab:fracsnana}.

\begin{table}[ht]
\begin{center}
 \small
\begin{tabular}{|c|c|c|} 
\hline
SNe Type & Li {\it et al.}(\%) & Input to \snana (\%) \\\hline \hline
Ib & 5.2 & 8.2  \\
Ic & 13.3 & 16.3  \\
Ibc-pec & 6.0 & 0  \\
II-P & 52.7 & 52.7  \\
II-L & 7.3 & 16.3  \\
IIb & 9.0 & 0  \\
IIn & 6.5 & 6.5  \\
\hline
\end{tabular}
\caption{We show a comparison of core collapse relative fractions from 
Li {\it et al.}~\cite{Li11} and those input to \snana.}\label{tab:fracsnana}
\end{center}
\end{table}

The uncertainties in the absolute brightnesses of SNcc are greater than the
uncertainties in the relative fractions.  The absolute 
brightnesses can dramatically affect the number of 
SNcc passing various selection cuts.
We have also made a change in the input parameters of \snana\ for the absolute brightnesses.  
Instead of the Richardson {\it et al.}~\cite{ric02}
brightnesses (corrected in an ad hoc way for Malmquist bias), 
we use the Li {\it et al.} brightnesses.  Table~\ref{tab:richvsli} compares the absolute
brightnesses and widths presented in Li {\it et al.} and Richardson {\it et al.}  One important aspect to note is 
that the Richardson {\it et al.} and Li {\it et al.} 
absolute brightnesses are in the B-band and R-band, respectively. 
Since the Li {\it et al.} measured widths are broader than 
those of Richardson {\it et al.}, we have added additional smearing to the input SNcc
templates in the \snana\ simulation.  A comparison of simulation-template magnitude offsets 
between Richardson {\it et al.} and Li {\it et al.} is presented in 
Tab.~\ref{tab:smear} (i.e. Simulated Template Peak Brightness = Measured Template 
Peak Brightness + Mag. Offset).  After the magnitude offset is 
added, an additional gaussian brightness smearing is 
applied with the one standard deviation values 
also presented in Tab.~\ref{tab:smear}.  The net effect of these changes
is to ensure that the SNcc template-brightness mean and RMS used in \snana\ match 
the results of Li {\it et al.}.

\begin{table}[ht]
\begin{center}
 \small
\begin{tabular}{|c|c|c|c|c|} 
\hline
SNe Type & $M_{B}$(Rich.) & $\sigma_{M_{B}}$(Rich.) & $M_{R}$(Li) & $\sigma_{M_{R}}$(Li)) \\\hline \hline
IP     & -14.40 $\pm$ 0.42 & 0.81 & -15.66 $\pm$ 0.16 & 1.23  \\
Ib$/$c & -16.72 $\pm$ 0.23 & 0.62 & -16.09 $\pm$ 0.23 & 1.24  \\
IIL    & -17.19 $\pm$ 0.15 & 0.47 & -17.44 $\pm$ 0.22 & 0.64  \\
IIn    & -17.78 $\pm$ 0.41 & 0.74 & -16.86 $\pm$ 0.59 & 1.61  \\
\hline
\end{tabular}
\caption{We present a comparison of means and RMS of the SNcc absolute-brightness
distributions from 
Richardson {\it et al.}~\cite{ric02} (B-band) 
and Li {\it et al.}~\cite{Li11} (R-band).}\label{tab:richvsli}
\end{center}
\end{table}

\begin{table}[ht]
\begin{center}
 \small
\begin{tabular}{|c|c|c|c|c|} 
\hline
SNe Type & Mag. Offset (Rich.) & Mag. Smearing (Rich.) & Mag. Offset (Li) & Mag. Smearing (Li)) \\\hline \hline
Ib$/$c     & 0.25  & 0.000 & 0.0 & 0.0  \\
Ib         & 0.00  & 0.000 & 0.5 & 0.1 \\
Ic         & 0.00  & 0.000 & 1.4 & 1.2 \\
IIn        & 0.00  & 0.742 & 1.0 & 1.5  \\
IIP        & 1.87  & 0.000 & 2.1 & 1.1  \\
IIL        & -0.30 & 0.469 & -0.80 & 0.6  \\
\hline
\end{tabular}
\caption{We show a comparison of core collapse \snana\ magnitude offsets and magnitude smearing from 
Richardson {\it et al.}~\cite{ric02} (B-band) and Li {\it et al.}~\cite{Li11}.(R-band).}\label{tab:smear}
\end{center}
\end{table}

\newpage

\section{Supplementary Figures}\label{apdx:supp}

In this appendix we present several supplementary figures;  each one is meant to complete
the set of SNR cuts shown in plots earlier in the paper.   The figure captions should be
self-explanatory. 

\begin{figure}[H]
\begin{center}
\includegraphics[width=0.49\columnwidth]{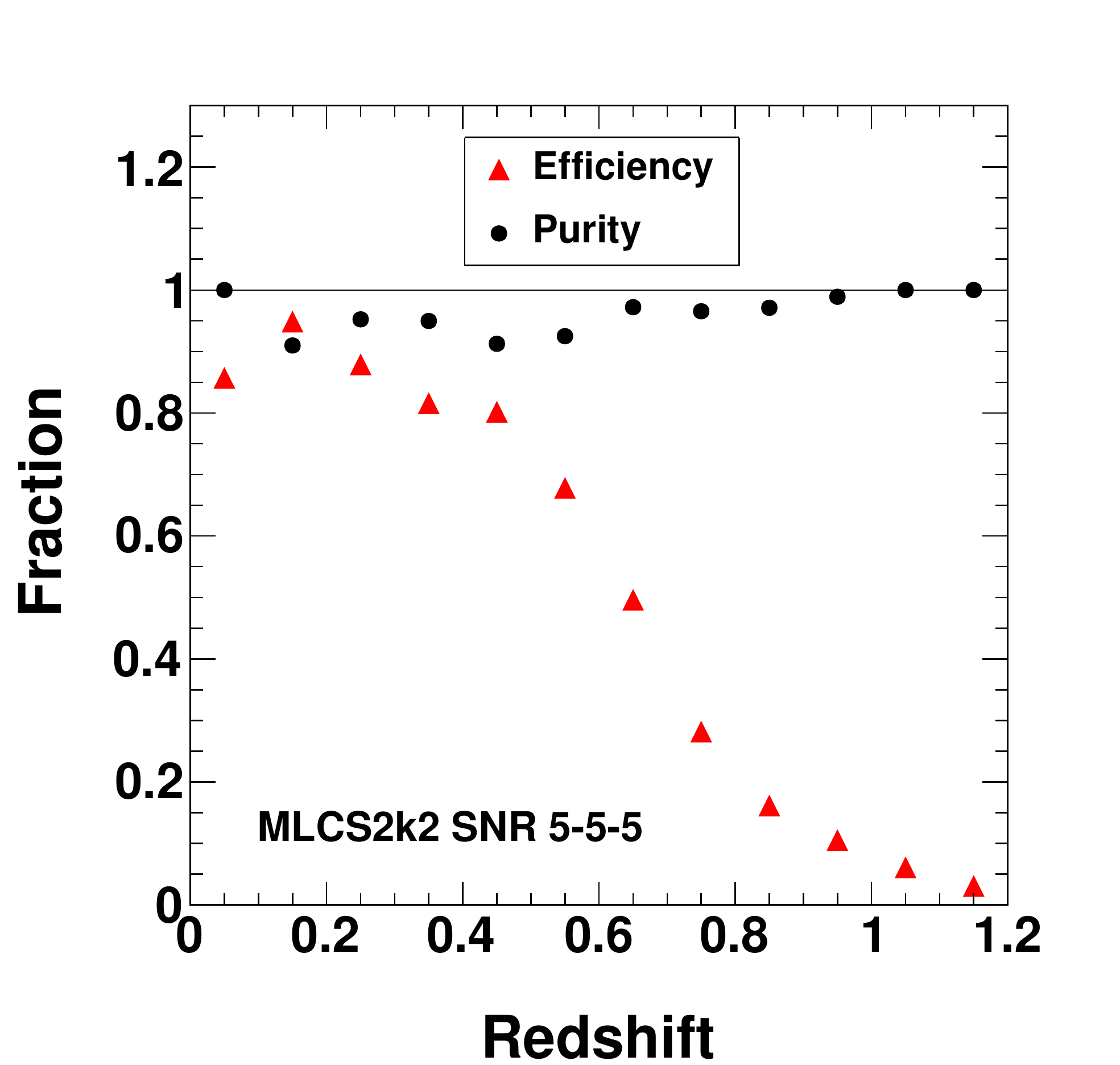}
\includegraphics[width=0.49\columnwidth]{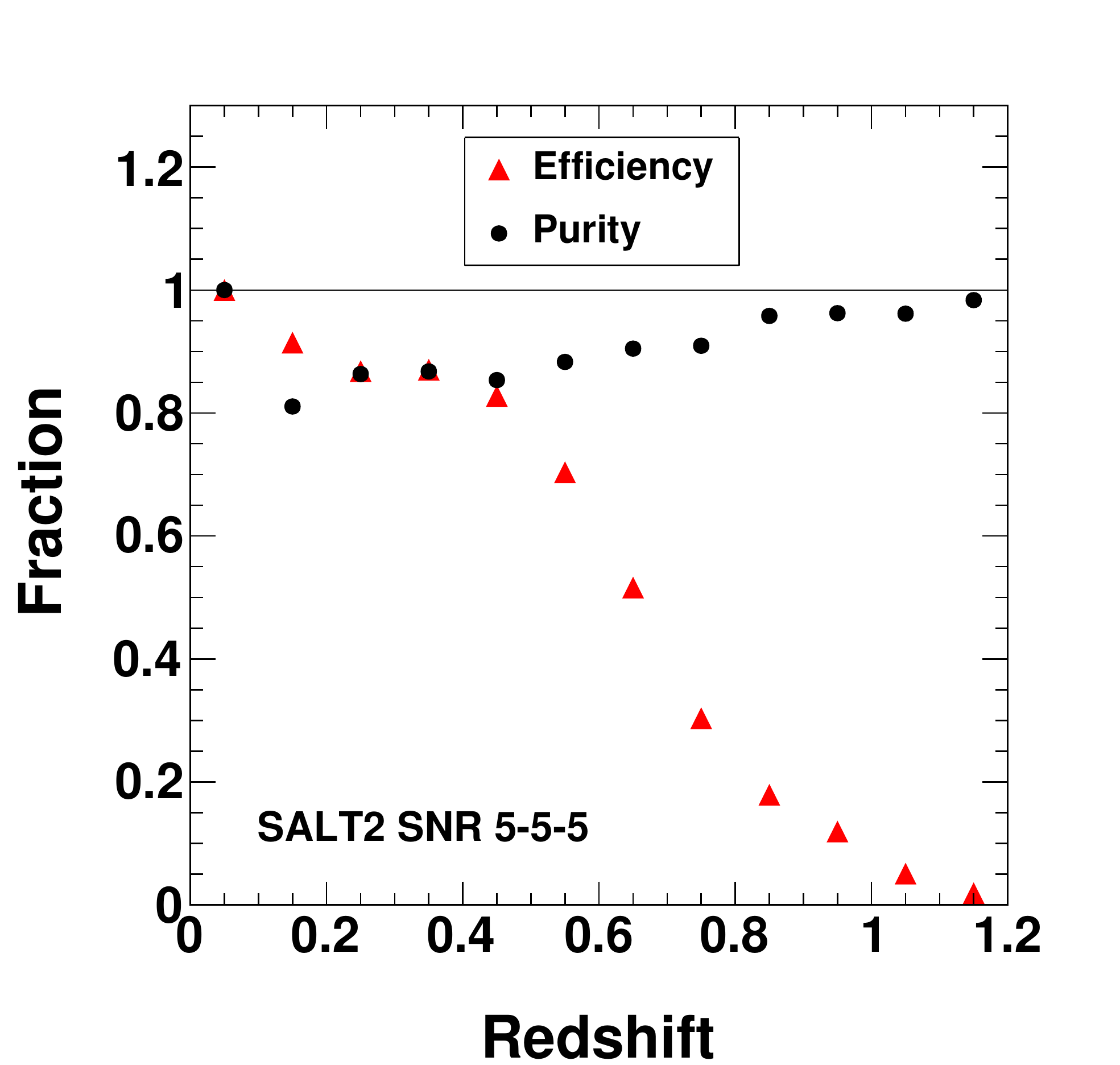}
\includegraphics[width=0.49\columnwidth]{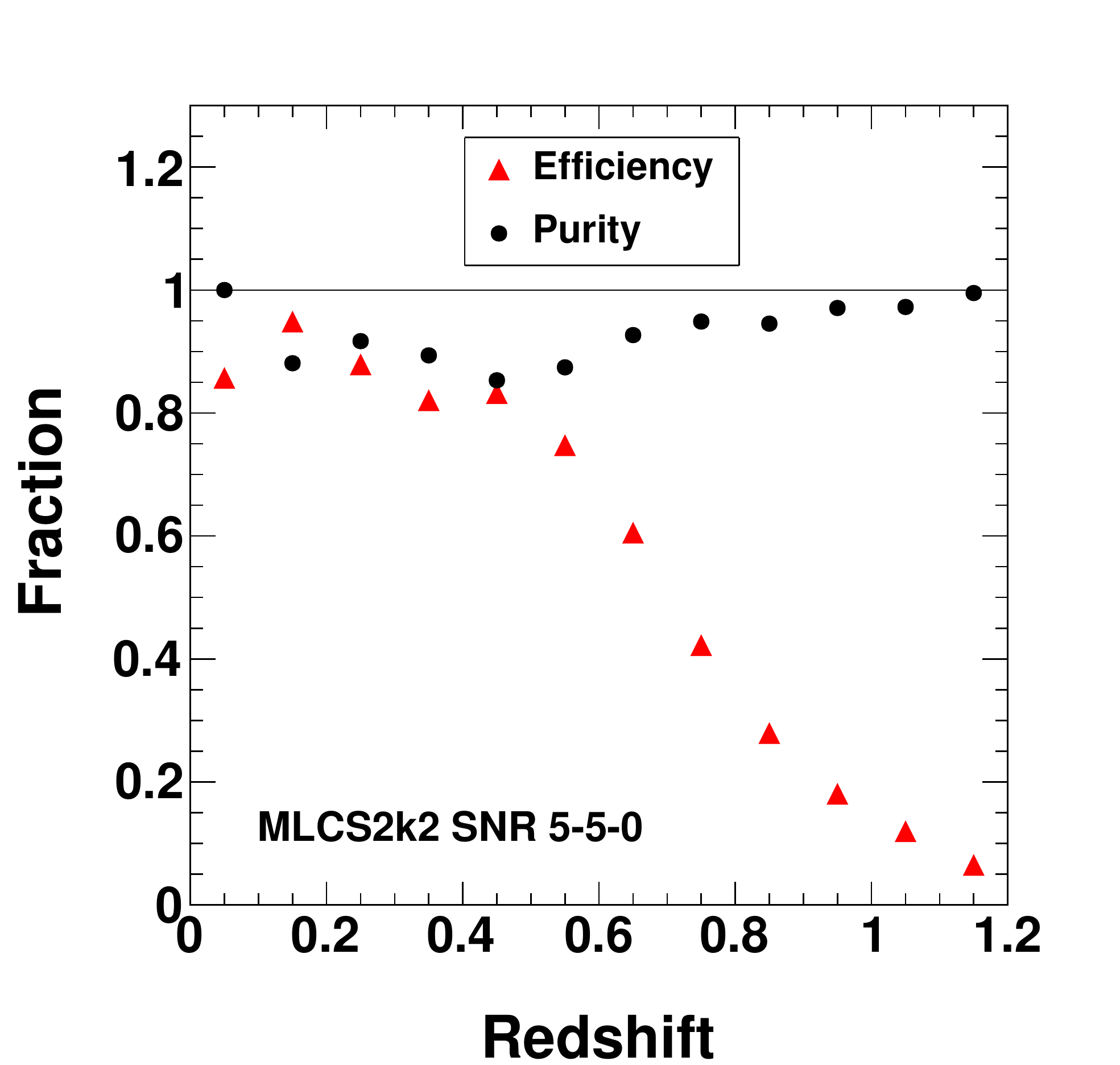}
\includegraphics[width=0.49\columnwidth]{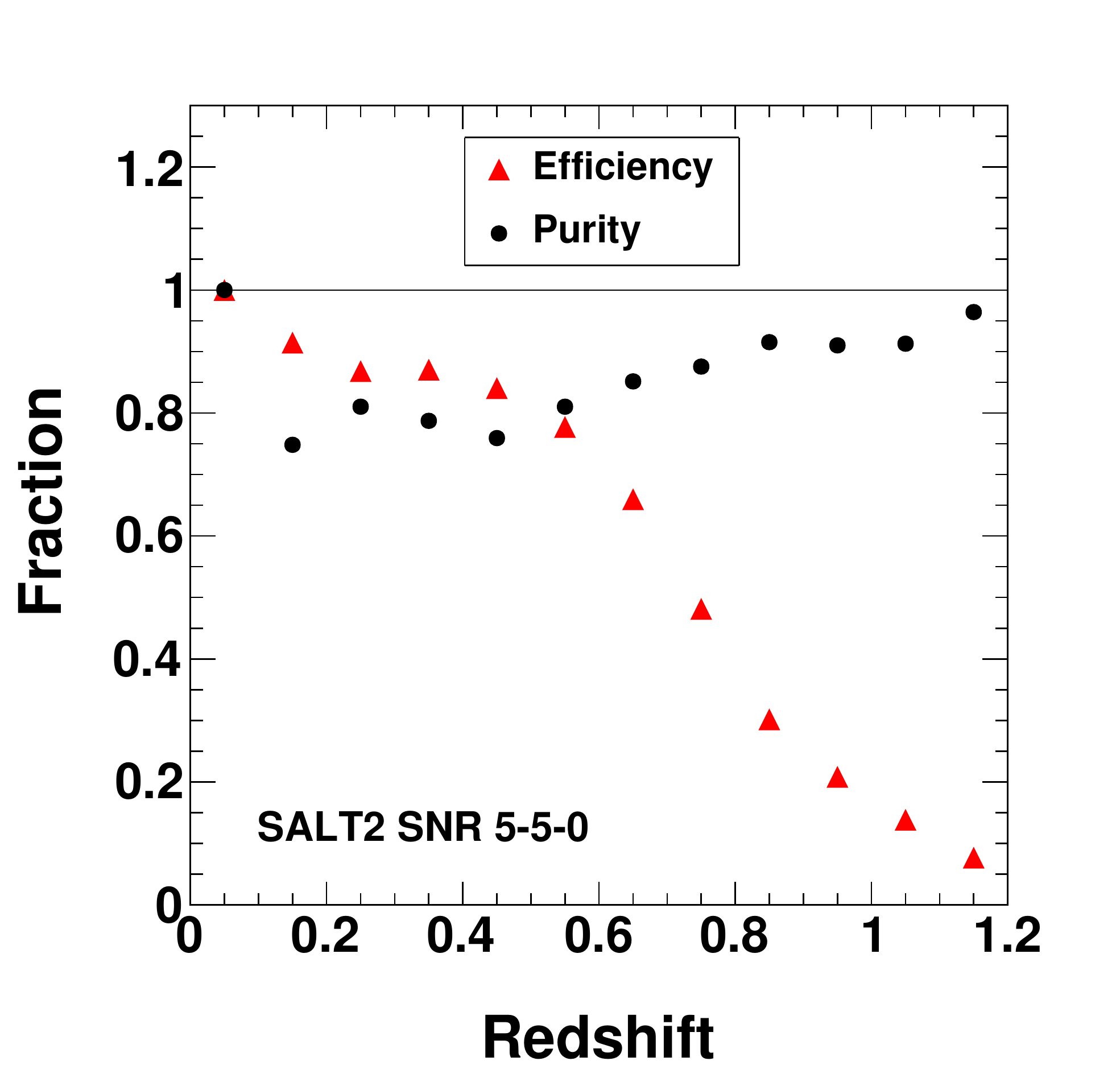}
\caption{As in Fig.~\ref{fig:purities}, the \mlcs\ purities and SNIa efficiencies ({\it left panels}) are plotted.
The definition of SNIa efficiency is the same as in Tab.~\ref{tab:purities}.
The top left panel is for the SNR cuts SNR-5-5-5,  while the bottom left panel is 
for the SNR cuts SNR-5-5-0.  
The corresponding \salt\ purities and SNIa efficiencies are shown in the right panels.
}\label{fig:purities2}
\end{center}
\end{figure}

\begin{figure}[H]
\begin{center}
\includegraphics[width=0.49\columnwidth]{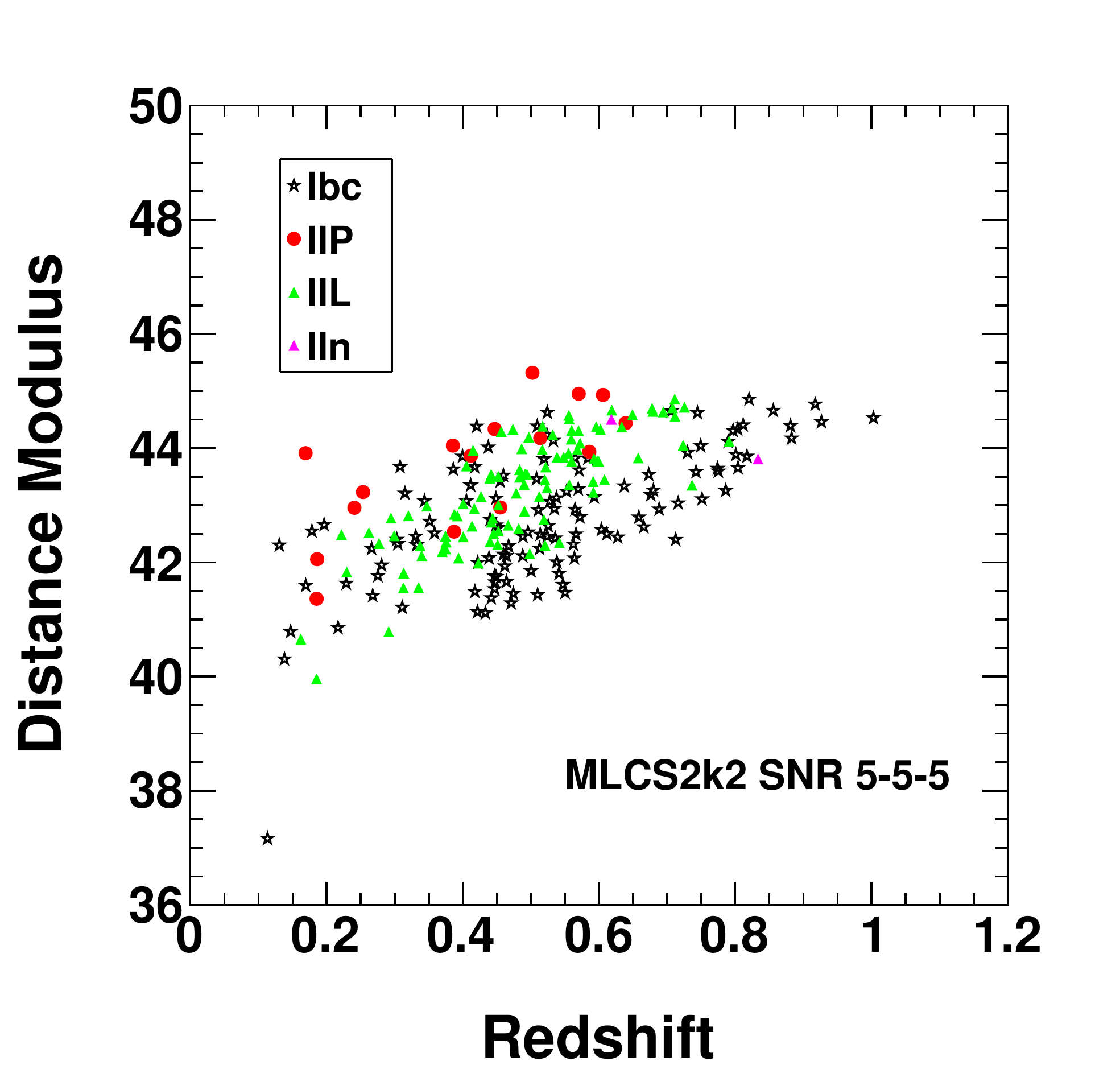}
\includegraphics[width=0.49\columnwidth]{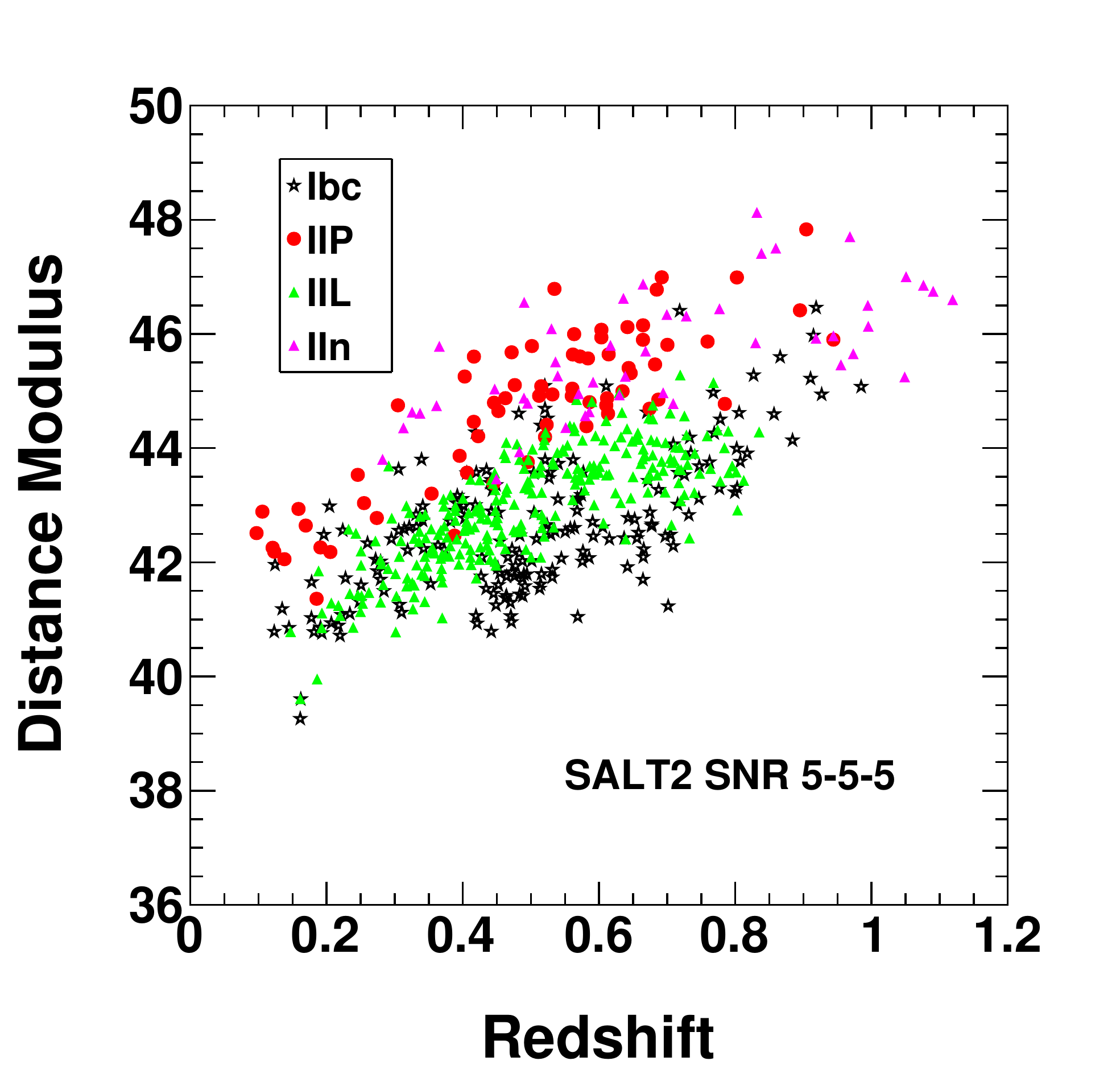}
\includegraphics[width=0.49\columnwidth]{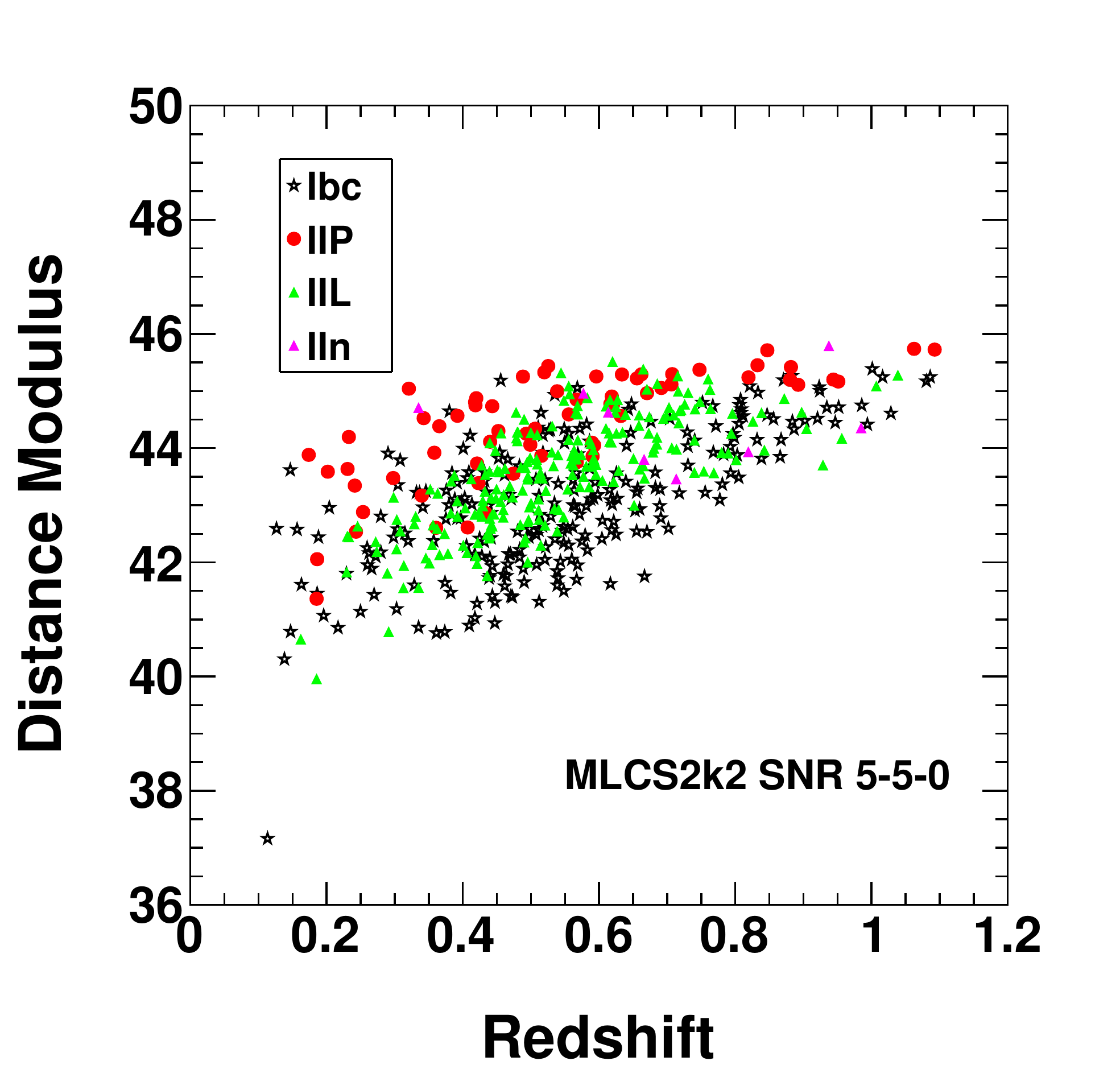}
\includegraphics[width=0.49\columnwidth]{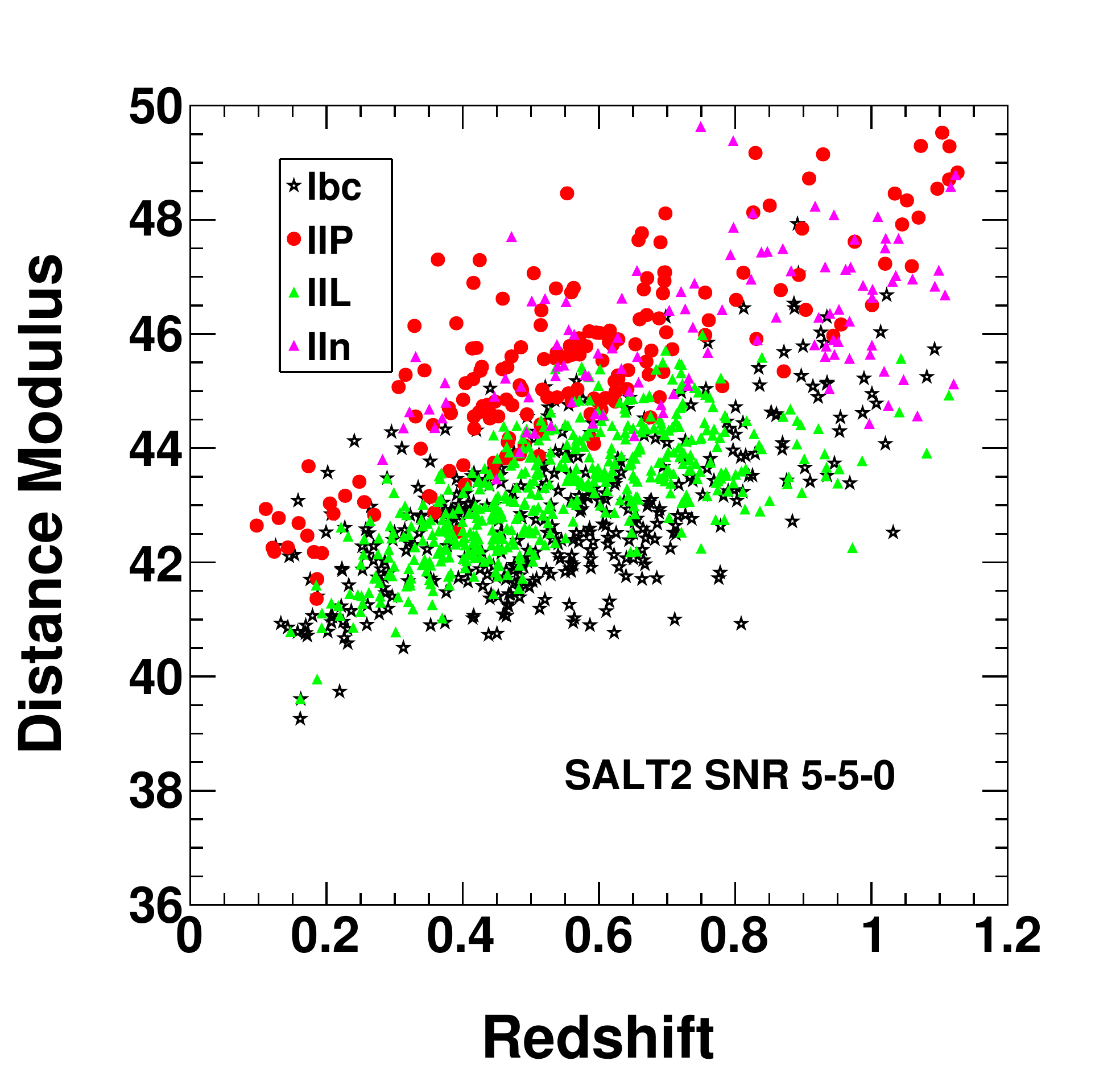}
\includegraphics[width=0.49\columnwidth]{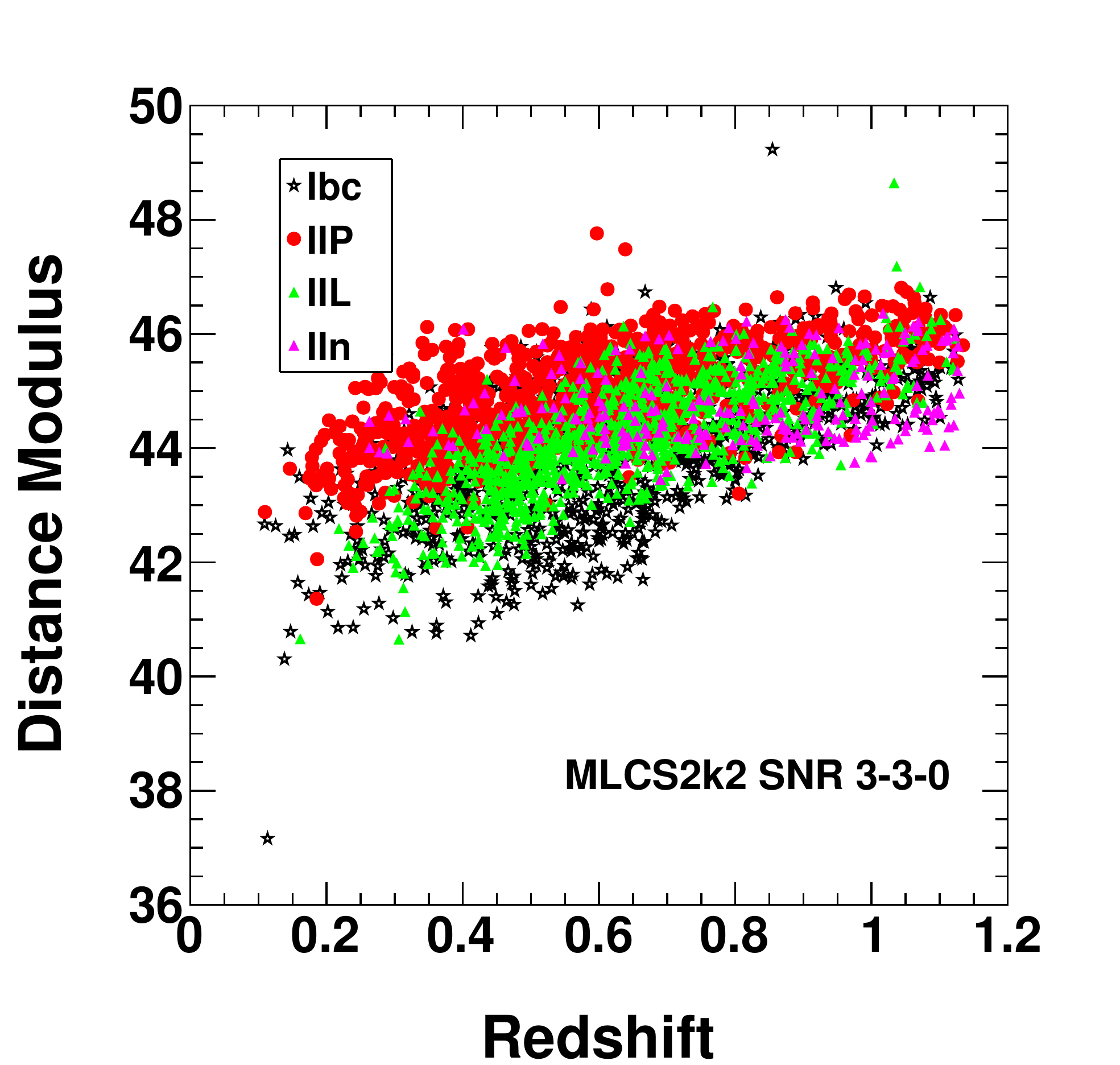}
\includegraphics[width=0.49\columnwidth]{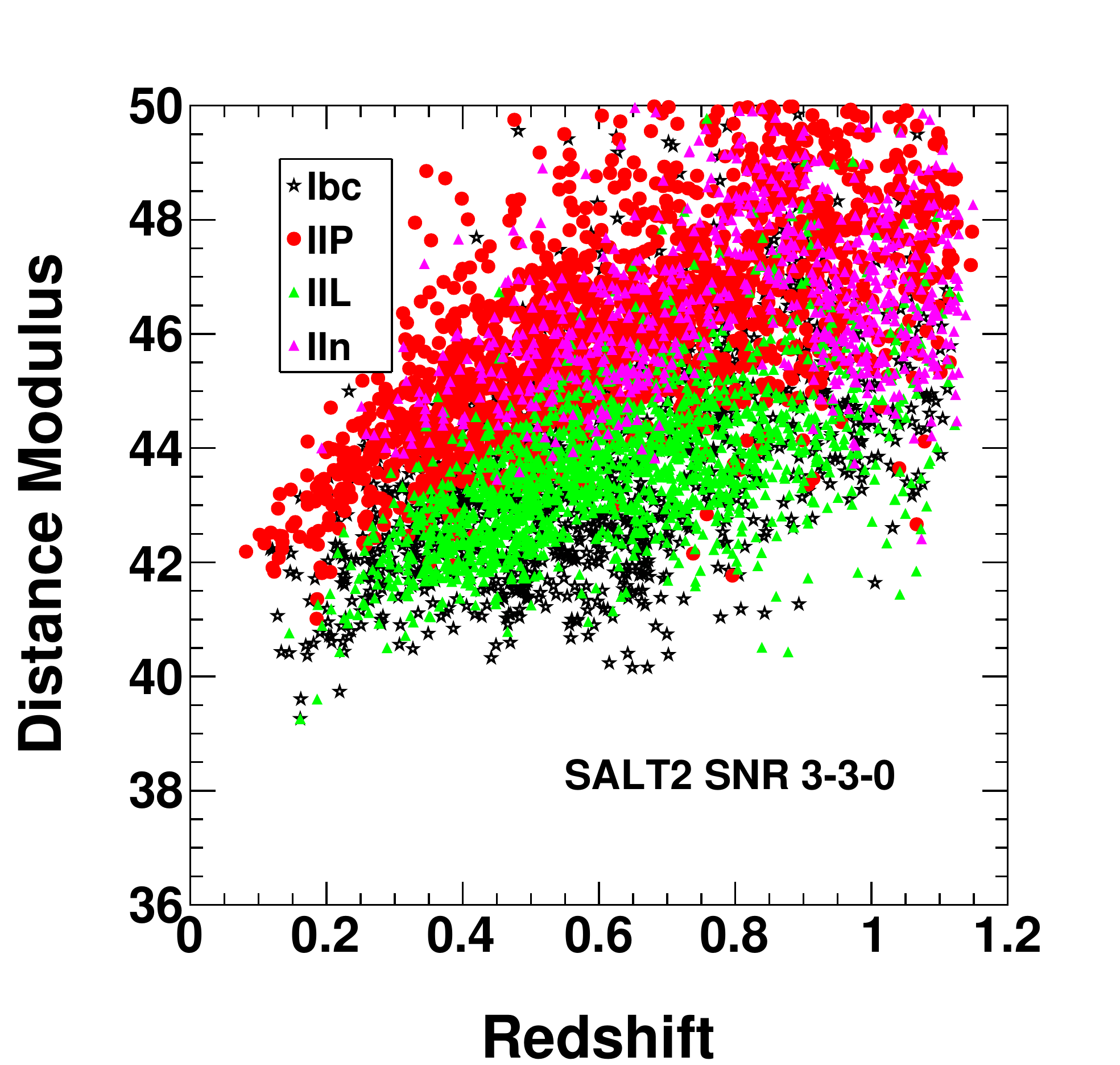}
\caption{The Hubble scatter is shown for the SNcc sample passing various
SNR cuts and a fit probability$>0.1$ cut in each case.  The left panels are
fit with the \mlcs\ model and the right panels are fit with the \salt\ model.
}\label{fig:types2}
\end{center}
\end{figure}

\begin{figure}[H]
\begin{center}
\includegraphics[width=0.49\columnwidth]{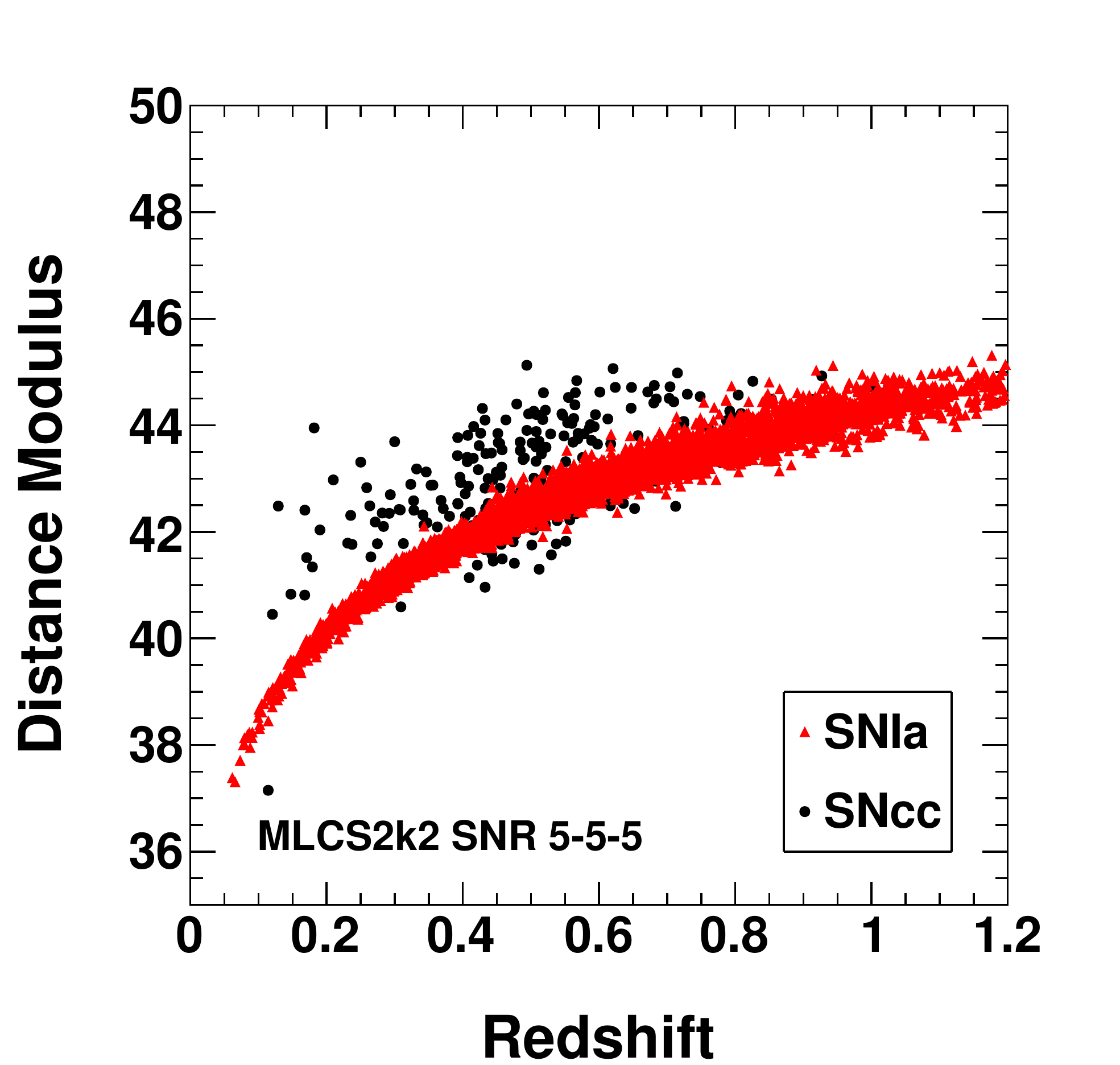}
\includegraphics[width=0.49\columnwidth]{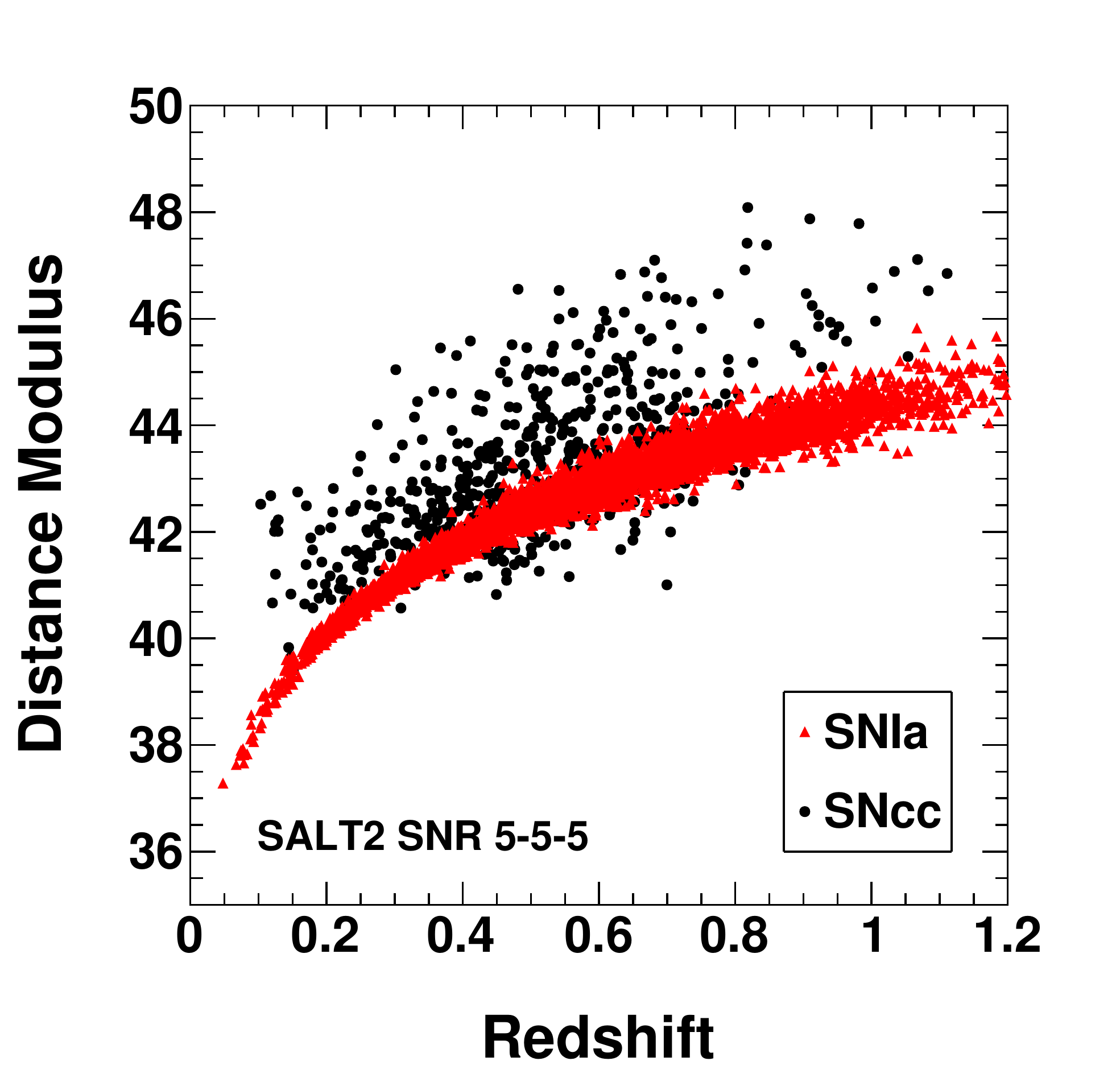}
\includegraphics[width=0.49\columnwidth]{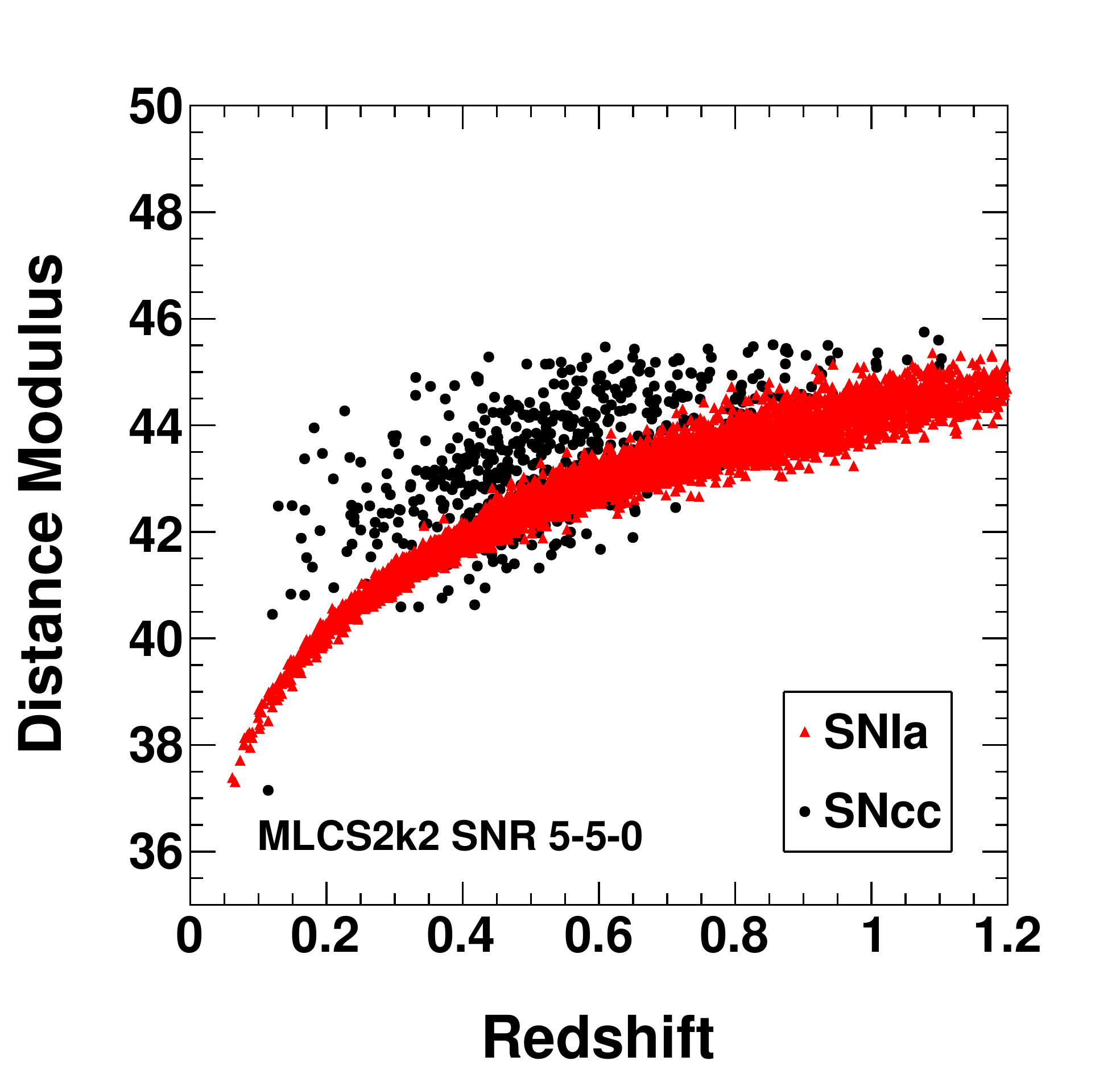}
\includegraphics[width=0.49\columnwidth]{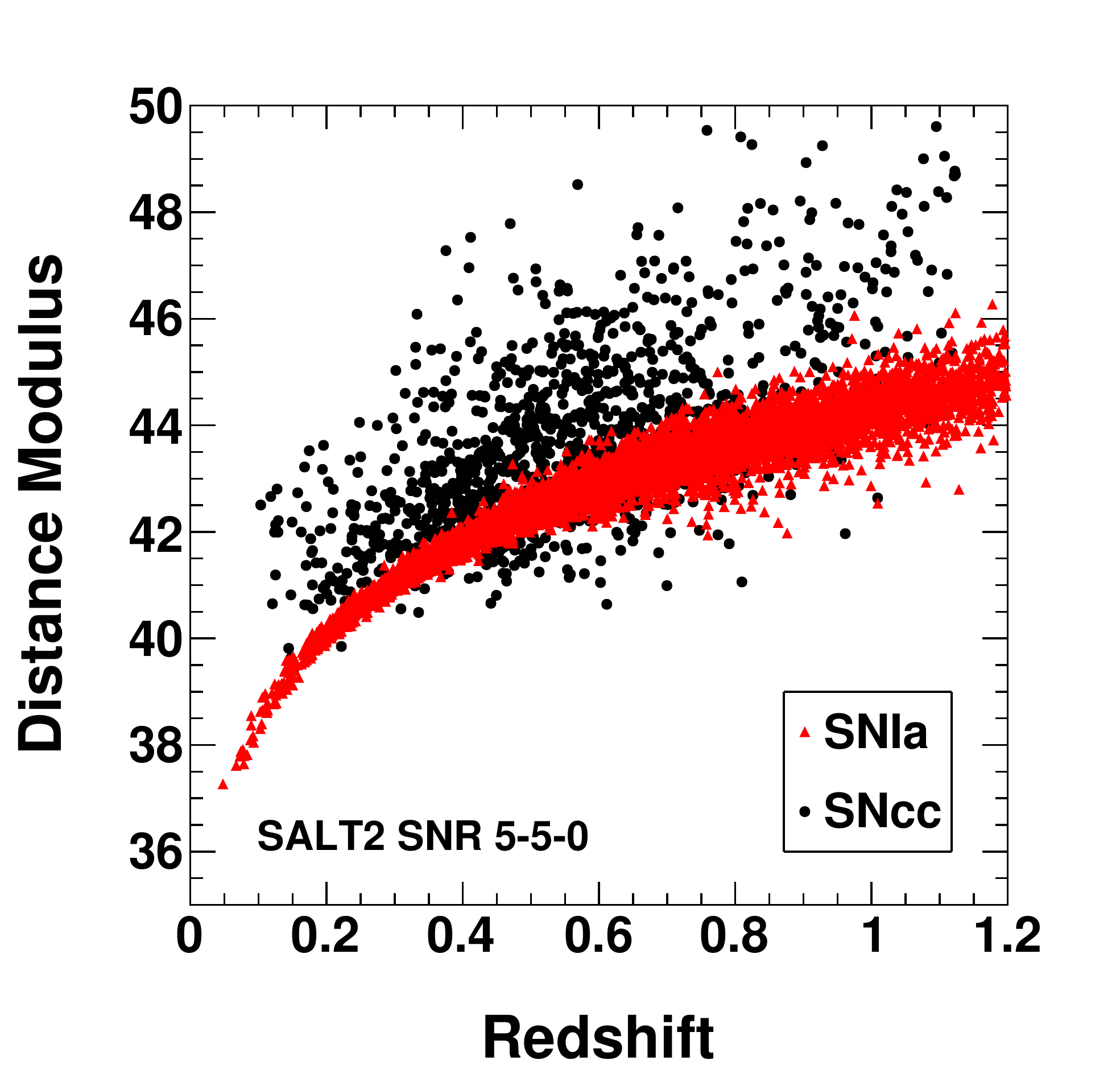}
\caption{As in Fig.~\ref{fig:hubblescatter}, we display four Hubble diagrams, 
for both SNIa and SNcc. The top panels are 
for the SNR-5-5-5 cuts,  while the lower panels are for the SNR-5-5-0 
cuts.  The left panels are with the MLCS2k2 light-curve fitter, while the right panels
are for the SALT2 light-curve fitter.}
\label{fig:hubblescatter2}
\end{center}
\end{figure}

\begin{figure}[H]
\begin{center}
\includegraphics[width=0.49\columnwidth]{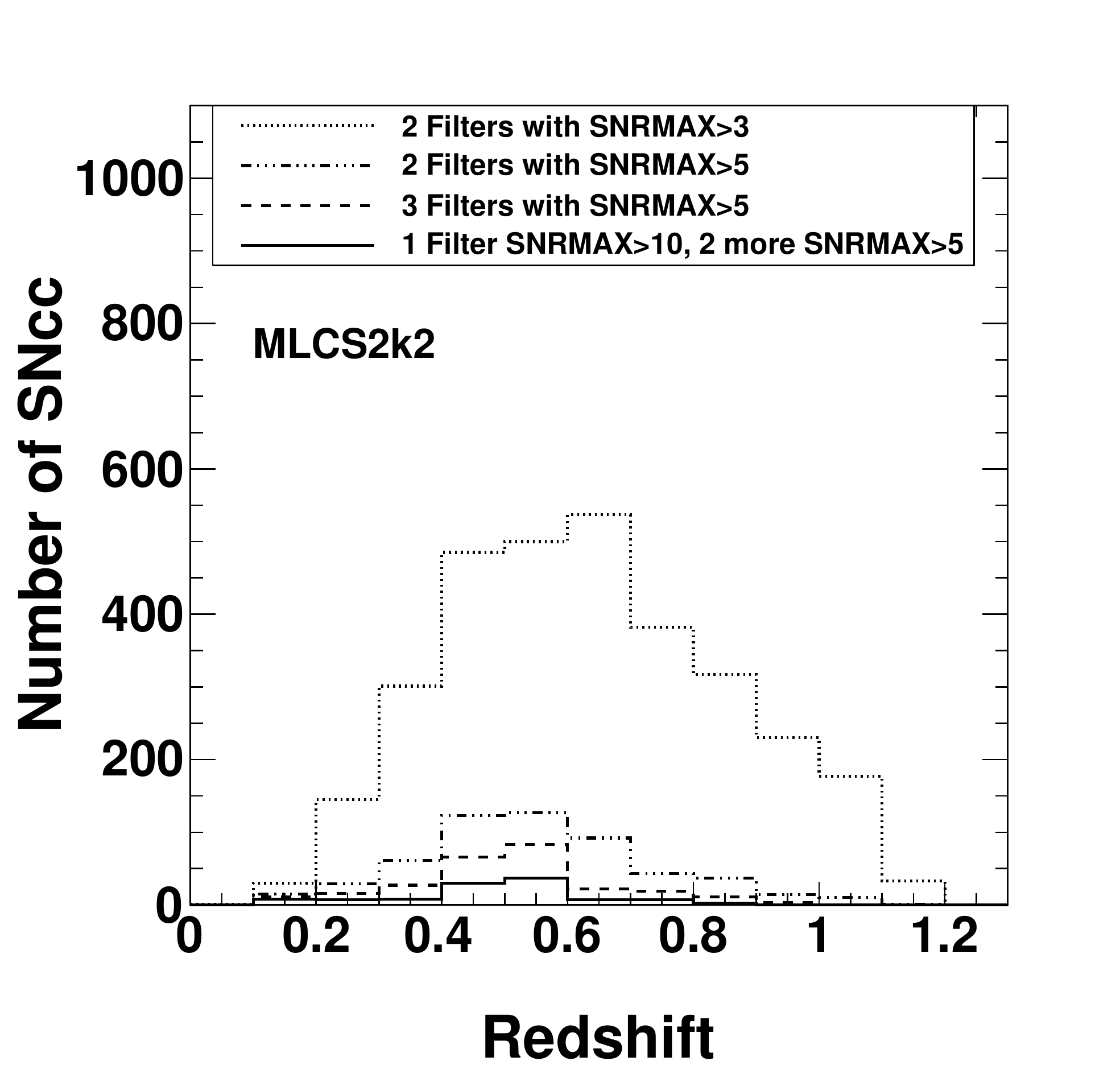}
\includegraphics[width=0.49\columnwidth]{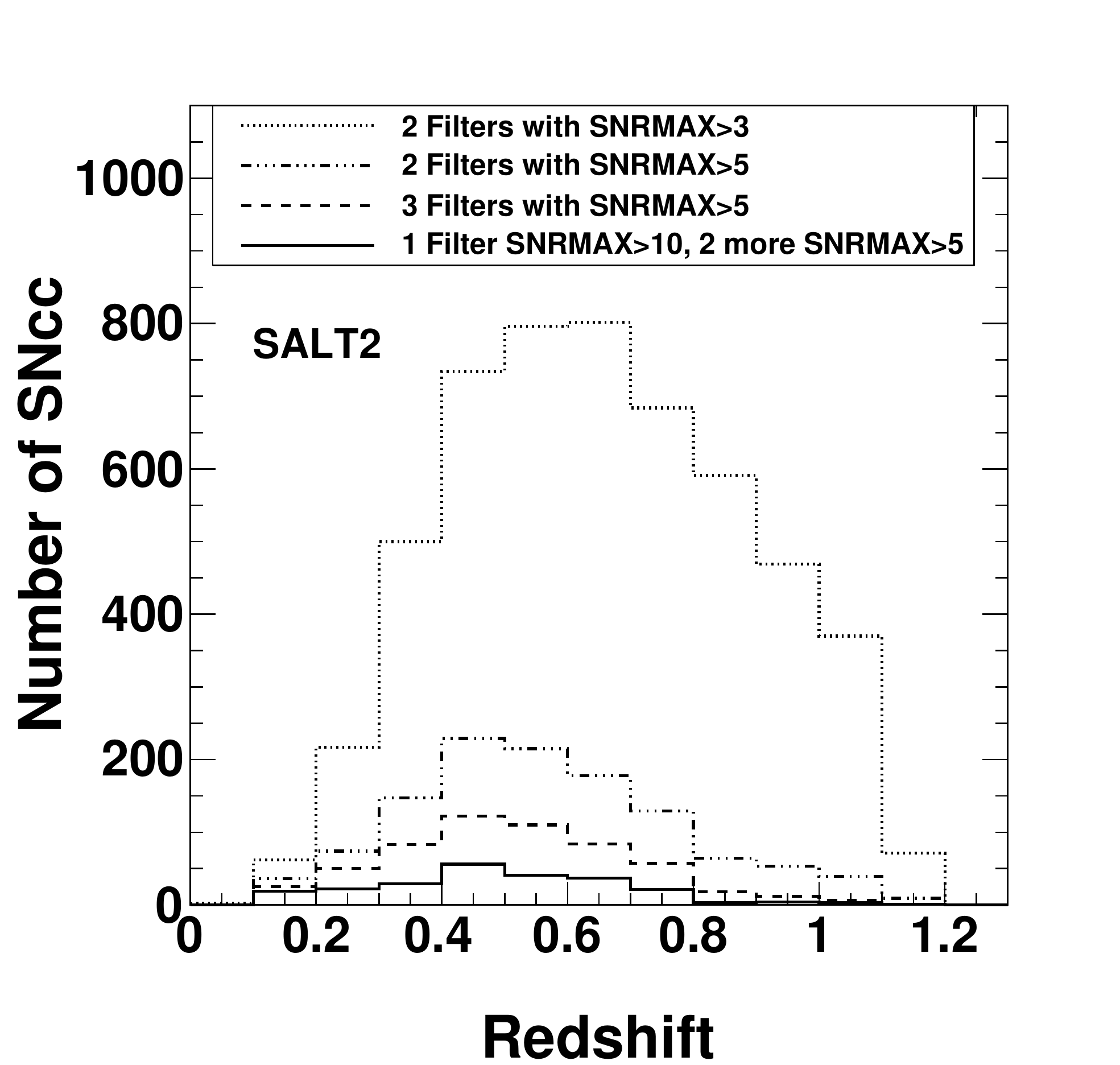}
\caption{As in Fig.~\ref{fig:zhists_and_types}, we display redshift distributions 
for each SNRMAX cut and a fit probability$>0.1$ cut,  this time for the core collapse 
samples (fit with \mlcs\ on the left and fit with \salt\ on the right). 
}\label{fig:zhistCC}
\end{center}
\end{figure}

\bibliographystyle{JHEP}

\end{document}